\documentclass[12pt]{article}

\usepackage[T1]{fontenc}
\usepackage{lmodern}
\usepackage{lscape}
\usepackage{comment}
\usepackage{ragged2e} 
\usepackage{siunitx} 
\usepackage{enumerate} 
\usepackage{pgfplotstable}
\usepackage{tikz,pgfplots}
\usepackage{microtype,amsmath}  %I added this so that you can use the align tool for equations!
\usepackage{amssymb}
\usepackage{wasysym} %This package allows you to put emojis in your paper!!!!
\usepackage{color}
\definecolor{brightmaroon}{rgb}{0.76, 0.13, 0.28}
\definecolor{blue(munsell)}{rgb}{0.0, 0.5, 0.69}
\usepackage{graphicx}
\graphicspath{{figures/}}  % Location of the graphics files (set up for graphics to be in PDF format)
\usepackage{subfigure}
\usepackage{scrextend}
\usepackage[bottom]{footmisc}

\usepackage{tabularx,booktabs,caption}
\usepackage[flushleft]{threeparttable}

\usetikzlibrary{bayesnet}
\usepackage[numbers, sort&compress, comma]{natbib}
\usepackage[colorlinks=true,linkcolor=blue(munsell),bookmarksopen,bookmarksnumbered, citecolor=blue(munsell),urlcolor=blue(munsell)]{hyperref}

\usepackage{algpseudocode,algorithm,algorithmicx}

\algrenewcommand\algorithmicrequire{\textbf{Precondition:}}
\algrenewcommand\algorithmicensure{\textbf{Postcondition:}}

\usepackage[euler]{textgreek}
\usepackage{forest,adjustbox}
\usetikzlibrary{arrows.meta, shapes.geometric, calc, shadows}
\colorlet{col3in}{gray!30}
\colorlet{col3out}{gray!40}
\colorlet{col5in}{blue!10}
\colorlet{col5out}{blue!20}
\colorlet{col6in}{white!20}
\colorlet{col6out}{white!30}
\colorlet{linecol}{gray!60}
\usepackage{geometry}
\pgfplotsset{compat=1.14}
\usepackage{authblk}

\begin{document}

%\title{A multivariate Bayesian learning approach for \\ anomaly detection within athletes' steroid profiles} %Title should be concise and to the point  

\title{A multivariate Bayesian learning approach for improved detection of doping in athletes using urinary steroid profiles}

% see Frimane2018 absract
\author[1]{Dimitra Eleftheriou}
\author[2]{Thomas Piper}
\author[2,3]{Mario Thevis}
\author[4]{Tereza Neocleous}

\affil[1]{\small{Leiden Academic Centre for Drug Research, Leiden University, The Netherlands}}
\affil[2]{\small{Center for Preventive Doping Research - Institute of Biochemistry, \vskip 0.05em
German Sport University Cologne, Cologne, Germany.}}
\affil[3]{\small{European Monitoring Center for Emerging Doping Agents (EuMoCEDA) \vskip 0.05em Cologne/Bonn, Germany}}
\affil[4]{\small{School of Mathematics and Statistics, University of Glasgow, United Kingdom}}

\date{}  % \today% This will automatically put today's date in the report

\maketitle  %this command makes the title

\begin{abstract}
Biomarker analysis of athletes' urinary steroid profiles is crucial for the success of anti-doping efforts. Current statistical analysis methods generate personalised limits for each athlete based on univariate modelling of longitudinal biomarker values from the urinary steroid profile. However, simultaneous modelling of multiple biomarkers has the potential to further enhance abnormality detection. In this study, we propose a multivariate Bayesian adaptive model for longitudinal data analysis, which extends the established single-biomarker model in forensic toxicology. The proposed approach employs Markov chain Monte Carlo sampling methods and addresses the scarcity of confirmed abnormal values through a one-class classification algorithm. By adapting decision boundaries as new measurements are obtained, the model provides robust and personalised detection thresholds for each athlete. We tested the proposed approach on a database of 229 athletes which includes longitudinal steroid profiles classified as normal, atypical, or confirmed abnormal. Our results demonstrate improved detection performance, highlighting the potential value of a multivariate approach in doping detection. %However, further testing on additional datasets is necessary to identify the optimal set of biomarkers for inclusion in the model, before the multivariate approach could be utilised more broadly in anti-doping monitoring.

\justify
\end{abstract}

\section{Introduction}
Doping has been widely discussed in recent years and remains a challenging topic in the athletic world. In competitive sports, prohibited substances such as anabolic androgenic steroids (AAS), refer to the drugs that are closely associated with the notion of doping \cite{kanayama2018history}. AAS are the most frequent detected class of substances in doping controls \cite{WADA2019testingfigures}. %that still gather a high number of adverse analytical findings (AAF) with regards to exogenous steroids and atypical passport findings (ATPF) after administration of pseudo-endogenous steroids, i.e. endogenous steroids that have been administered exogenously \cite{WADA2019testingfigures}. 
%(please add here the testing figures report of WADA as citation: https://www.wada-ama.org/sites/default/files/resources/files/2019_anti-doping_testing_figures_en.pdf)  
In order to detect the administration of endogenous steroids, i.e. steroids that are produced inside the body such as testosterone,  the steroidal module of the Athlete Biological Passport (ABP) was developed \cite{sottas2010endogenous}. The steroidal module is used to denote a follow-up which is the recording of the concentration of endogenous steroids and their ratios in urine over time. In order to establish the ABP, the World Anti-Doping Agency (WADA) provided harmonised and robust analytical methods for the ``steroid profile'' which according to their technical document (TD) \cite{WADA2021techdoc} is composed of the following endogenous anabolic androgenic steroids (EAAS): testosterone (T), epitestosterone (E), androsterone (A), etiocholanolone (Etio), 5\textalpha-androstane-3\textalpha, 17\textbeta-diol (A5), 5\textbeta-androstane-3\textalpha, 17\textbeta-diol (B5), %dehydroepiandrosterone (DHEA), dihydrotestosterone (DHT), 
as well as the concentration ratios T/E, A/T, A/Etio, A5/B5 and A5/E. As the intra-individual variation of all these markers is much lower than the inter-individual variation in the population of athletes, individual longitudinal monitoring as applied by the ABP increases the sensitivity for the detection of illicit AAS administration. For each athlete an individual reference range is specified for each biomarker as samples are added to the ABP over time. If a new steroid profile enters into the individual ABP and values fall beyond established thresholds, the ABP alerts %The fact that athletes have their own metabolism and different responses after a drug intake creates significant inter-individual variation. The method of ABP steroid profiling fights against doping, overcoming the difficulty of population cut-offs. Subsequently, if these biomarker levels change significantly within the steroid profile of an athlete, it alerts 
athlete passport management units (APMUs) that anomalies have been detected that require closer examination \cite{sottas2010endogenous, sottas2011athlete, vernec2014athlete, WADA2021TDAPMU}. %(please add here another citation on the technical document of WADA dealing with APMUs: https://www.wada-ama.org/sites/default/files/resources/files/td2021apmu_final_eng_0.pdf) 
Employing this approach, the ABP mostly aids in revealing the direct and indirect effects of doping with anabolic agents on the individual steroid profile rather than detecting the prohibited substance itself \cite{kuuranne2014confounding, piper2021current}. Therefore, further steps during the closer examination encompass steroid profile confirmation, detection of confounding factors and isotope ratio mass spectrometry (IRMS) based investigations. 
\justify
By far the most commonly used screening tool in %forensic toxicology for doping detection 
sports drug testing over the last years uses the T/E ratio, as it is considered a stable marker ratio within an athlete's steroid profile, sensitive to the administration of T itself and T-prohormones \cite{mareck2008factors}. Measuring only the urinary T concentrations has proved inadequate due to the small ratio of intra- to inter-individual variability in the urinary steroid concentrations caused by various factors \cite{harris1974effects, brooks1979detection,donike1983nachweis, sottas2006statistical, mareck2008factors}. Therefore, monitoring the steroid profile at individual level is very important, mostly because the reference values based on the population do not always have the sensitivity to track whether anabolic drugs have been administered \cite{sottas2008population, van2010reference}. Current approaches receive new measurements of a single biomarker or ratio and, under a Bayesian framework, progressively adapt population-derived limits, when there are no recorded measurements, to individual normal ranges as the number of measurements increases \cite{sottas2007bayesian}. Multivariate statistical approaches have also been proposed for this purpose, which are able to combine population information with individual longitudinal monitoring of multiple biomarkers \cite{brown2001bayesian, alladio2016application, amante2019multivariate}. However, multivariate statistical methods in a Bayesian adaptive framework have not yet been attempted.
\justify
This research focuses on the development of an improved statistical model for classifying athletes' urine samples into \textit{suspicious} and \textit{non-suspicious} classes. The classification technique is based on comparisons of sequential measurements of a set of biological variables against previous recordings. For this purpose, we gradually move from the univariate to a multivariate statistical model which considers prior information on inter- and intra-individual variations, and on potential correlation between the EAAS markers. A fully validated method using GC-MS analysis and 
fulfilling all requirements as per TD EAAS \cite{mareck2008factors, WADA2021techdoc} determined the six markers and the five ratios which compose the urinary steroid profile of the ABP of 229 athletes. Out of them, 100 were athletes whose samples are considered to be \textit{normal} (i.e. none of their samples was found beyond the individual limits), 100 were athletes with one or more extreme samples classified as \textit{atypical} (i.e. samples beyond the individual limits but in none of these a doping offence was detectable), and 29 confirmed dopers, each marked with at least one atypical sample in their longitudinal steroid profile and confirmed to be a real doping offence employing IRMS according to WADA regulation \cite{WADATD2021IRMS}. The latter samples are classified as \textit{abnormal}.  %Out of them, 100 were non-abusers whose samples are considered to be \textit{normal}, 100 were non-abusers with some extreme samples classified as \textit{atypical}, and 29 athletes, who volunteered to receive T-gel, each had at least one confirmed \textit{abnormal} sample in their steroid profile employing to IRMS according to WADA regulation \cite{WADATD2021IRMS}.  
%\textbf{ADD A SENTENCE EXPLAINING WHAT ABNORMAL MEANS: CONTROLLED ADMINISTRATION OF TESTOSTERONE OR SOMETHING ELSE? SIMILARLY WITH ATYPICAL} 
These EAAS concentrations were longitudinally sampled from professional athletes, whereas cross-sectional measurements of the same markers and ratios from a baseline population of 164 %74+92-2=166-2  added two populations of sport students and employees of the Sports University here gathered in 2008 and 2011
healthy and non-doped volunteers were used to extract prior information for the normal range of steroid concentrations. It is worth noting that recordings from athletes with normal concentration values are more readily available than from doped athletes. Since the imbalance in size between the two classes is unavoidable, traditional classification methods may not work effectively as they are biased towards the predominant class. Hence, we classify athletes' samples based on a one-class classification algorithm. The AAS concentrations from non-doped athletes define the ``target'' class for which adaptive decision boundaries are constructed to separate them from abnormal data (also known as outliers or anomalies). 
\justify
The remaining sections are structured as follows. In Section 2 we introduce two models; a univariate and a multivariate Bayesian adaptive model. The former can be applied to individual biomarkers or ratios within the steroid profile, while the latter is designed to analyse multiple biomarkers and/or ratios simultaneously. Section 3 describes the one-class classification method using the highest posterior predictive density (HPD). In Section 4 we present summaries of real athletes' data and the HPD estimation on both models. The inference and computations in the analysis %for the out-of-sample predictive distribution 
relied on either sampling from a known joint posterior distribution or sampling from a non-closed form distribution employing Markov chain Monte Carlo (MCMC) methods, depending on the model being used. The computer code for the developed models is written in the statistical language {\fontfamily{cmtt}\selectfont R} 4.2.2 \cite{RProject}. We compare the two models with each other but also assess them against a simpler generalised linear mixed effects model (GLMM) and the univariate Bayesian approach introduced by \citet{sottas2007bayesian}, exclusively designed for the analysis of the T/E ratio. Furthermore, we conduct a comparison with the univariate Bayesian model applied to the Euclidean distance score as outlined in the work of \citet{de2023new}. %\textbf{ADD HERE THAT WE COMPARE ALSO TO THE UNIVARIATE BAYESIAN MODEL OF SOTTAS APPLIED TO THE SCORE, AS PER ACA2023, Figueiredo et al, WHICH NEEDS TO BE ADDED TO THE REFERENCES. POSSIBLY REMOVE THE REFERENCE TO THE GLMM UNLESS YOU'D LIKE TO ADD SOME MORE INFORMATION ON IT SOMEWHERE.}
Section 5 discusses the main findings and directions for further research. 

%We compare the multivariate Bayesian model with other statistical methods previously used for the same purpose. 
%We also compare the developed model with the univariate Bayesian approach introduced by \citet{sottas2007bayesian} in 2007, which uses a joint prior distribution that has been designed exclusively for the analysis of the T/E ratio. 
%We present a univariate hierarchical Bayesian model that can be applied to any biomarker or ratio of the steroid profile by assuming Gaussian distributions and conditionally conjugate priors. 
%Inference and computations, such as the out-of-sample predictive distribution, were based on both techniques, i.e. sampling from the known joint posterior distributions and using Markov chain Monte Carlo (MCMC) sampling methods, depending on the model. The computer code for the developed Bayesian model is written in the statistical language {\fontfamily{cmtt}\selectfont R} 4.1.1. %A user-friendly software tool was developed to implement the methodology, so that it can be used by practitioners as widely as possible. The latest version of the application can be found at \href{https://dimitraelegla.shinyapps.io/doping_shiny_app/}{BioScan App}. %?

% cross validation etc comment
% eleonora 2019
% The psychological stress also proved to influence the steroids [30-32] secretion from HPA and gonads. The relationship between the stress due to competition and the increase of testosterone was demonstrated by Guezennec et al., who monitored the testosterone levels in plasma before and after pistol shooting.

\section{Models}

\subsection{The univariate Bayesian model}
\label{univarmodel} 

In this section we introduce an expanded version of the univariate Bayesian model for the T/E ratio proposed by \citet{sottas2007bayesian} to a univariate Bayesian hierarchical model for any steroidal component or ratio of the ABP \cite{eleftheriou2022bayesian}. Under a Bayesian framework, the model receives new measurements and progressively adapts population-derived limits, when the number of measurements $n$ is zero, to individual normal ranges, when $n$ is large. %Throughout the paper, all vector quantities are denoted by bold-faced characters. 
\justify
Let $\boldsymbol{y}$ $=(y_{1},...,y_{n})$ represent the vector with the $n$ log-transformed recorded values of the biomarker $Y$ collected from the same athlete. It is important to mention that the period among two sequential samples of an athlete is long enough for them to be considered independent. Hence, we assume the logarithm of EAAS values to be a vector of $n$ independent and identically distributed draws from a Gaussian distribution with mean $\mu$ and variance $\sigma^2$, that is 

\vspace{-0.2cm} 
\begin{equation}
\label{equ:gaussian}
Y | \mu, \sigma^2 %\stackrel{iid}
{\sim} \mathcal{N}(\mu,\sigma^{2}).
\end{equation} 

%\textbf{DO YOU MEAN $Y_i, i=1,\dots, n$ ARE IID GAUSSIAN WITH MEAN $\mu$ AND VARIANCE $\sigma^2$ EACH?}
\justify
We focus on modelling the logarithm of these EAAS concentrations due to the fact that there are physical constraints on the measurement values (i.e. all markers are positive) and taking the logarithm allows us to use the Gaussian distribution in our models. To describe prior knowledge about the unknown parameters, i.e. the mean $\mu$ and the precision $\tau = 1 / \sigma^2$, we specify the joint prior distribution as the product of a conditional and a marginal distribution expressed as $p(\mu,\tau)= p(\mu| \tau) p(\tau)$. Given our limited prior information on the parameters of the model regarding the six available biomarkers and their five ratios with the exception of the T/E ratio, we propose specifying weakly informative conditionally conjugate priors on these model parameters. Therefore, a Gaussian distribution is assigned to the mean conditional on the precision as 
\begin{equation}
\label{priormean}
\mu|\tau \sim \mathcal{N}(\mu_0,1/ (\kappa_0 \tau)),
\end{equation}
\justify
 with hyper-parameters $\mu_0$ (prior mean) and $\kappa_0$ (prior sample size, which determines the tightness of the prior). The inverse of the variance is assumed to follow a conjugate Gamma prior distribution expressed as
 \vspace{-0.2cm}
\begin{equation}
\label{priorprecision}
\tau \sim \mathcal{G}a ( \alpha_0, \beta_0 ),
\end{equation}
\justify
where $\alpha_0$ and $\beta_0$ determine the \textit{shape} and \textit{rate} hyper-parameters, respectively. The T/E ratio is excluded from the semi-informative prior setting because there is adequate population information regarding its characteristics given by \citet{sottas2007bayesian}. However, we present the analysis results of T/E using both informative and weakly informative priors for comparison purposes. As a conjugate family, the joint posterior distribution for the pair of parameters $(\mu,\tau)$ is also a Gaussian-Gamma distribution 

\vspace{-0.2cm}
\begin{equation}
\label{posteriorGG}
\mu,\tau| \boldsymbol{y} \sim \mathcal{NG}a(\mu_n, \kappa_n, \alpha_n, \beta_n), 
\end{equation}
\justify
where the index $n$ on the hyper-parameters indicates the updated values after seeing the $n$ observations from the sample data, that is

\vspace{-0.65cm}
\begin{equation*}
\begin{split}
\label{updates}
\kappa_n & = \kappa_0 + n, \hspace{0.5cm}
\mu_n  = \frac{\kappa_0\mu_0 + n\bar{y}}{\kappa_n}, \\
\alpha_n & = \alpha_0 + n/2, \hspace{0.5cm}
\beta_n = \beta_0 + \frac{1}{2} \sum_{i=1}^{n}(y_i - \bar{y})^2 + \frac{\kappa_0n}{2\kappa_n}(\bar{y}-\mu_0)^2.
\end{split}
\end{equation*}

\subsection{Proposed multivariate Bayesian adaptive model}
\label{multivmixed}
\begin{figure}[h]
  \begin{center}
  \begin{tikzpicture}

  % Layout the variables
  \matrix[row sep=0.5cm, column sep=1.3cm] (MGME)
  { %
    \node[latent,xshift=-1cm] (muj)   {$\boldsymbol{\mu}$} ; & %
    \node[latent] (Sj)    {$\boldsymbol{b_j}$} ; & %
    \node[latent,xshift=2cm] (beta0)    {$\Omega_e$} ; & %
    \\
    \\
    &  %
    \node[obs,rectangle] (Xi)   {$\boldsymbol{y_{ij}}$} ;
    \\
  };

  % Remaining factors
  \factor[above=of muj]     {muj-f}      {left:Gaussian} {} {}; %
  \factor[above=of Sj]      {Sj-f}      {left:Gaussian} {} {}; %
  \factor[above=of beta0]    {beta0-f}    {right:Wishart} {} {};
  \factor[above=of Xi] {Xi-f} {left:Gaussian} {} {}; % 

  % Hyperparameters
  \node[latent,above=of muj,xshift=0.5cm] (S0) {$\Omega_{\mu}$} ; %& %
  \node[latent,above=of Sj] (W0)    {$\Omega_b$} ; 
  \node[latent,rectangle, above=of S0, xshift=-0.5cm] (alambda1)
  {$S_{\mu}$}; %
  \node[latent,rectangle, above=of S0, xshift=0.5cm] (alambda2)
  {$d_{\mu}$}; %
  \node[latent,rectangle, above=of beta0,xshift=-0.5cm]     (d1)    {$S_e$}; %
  \node[latent,rectangle, above=of beta0,xshift=0.5cm]     (d2)    {$d_e$}; %
  \node[latent,rectangle,above=of W0,xshift=-0.5cm] (W01)    {$S_b$} ;  %
  \node[latent,rectangle,above=of W0,xshift=0.5cm] (W02)    {$d_b$} ;  %
 \node[latent, rectangle, above=of muj, xshift=-0.5cm] (mu0)  {$\boldsymbol{\mu_0}$} ; %& %

  \factor[above=of S0] {S0-f} {left:Wishart} {} {}; %
  \factor[above=of W0] {W0-f} {left:Wishart} {} {}; % 

  % Factor connections
  \factoredge {mu0}             {muj-f}      {muj} ; %
  \factoredge {beta0}            {Xi-f}      {Xi} ; %
  \factoredge {W0}              {Sj-f}      {Sj} ; %
  \factoredge {alambda1,alambda2} {S0-f} {S0} ; %
  \factoredge {d1}              {beta0-f}    {beta0} ; %
  \factoredge {d2}             {beta0-f}      {beta0} ; %
  \factoredge {W01}             {W0-f}      {W0} ; %
  \factoredge {W02}             {W0-f}      {W0} ; %
  \factoredge {S0}           {muj-f}     {muj}; %
  \factoredge {muj}          {Xi-f}      {Xi} ; %
  \factoredge {Sj}           {Xi-f}      {Xi} ; %

  % Plates
  \plate {P1} { %
%    (Ci)(Ci-f)(Ci-f-caption) %
    (Xi)(Xi-f)(Xi-f-caption) %
  } {$i=1,\dots,n_j$} ;

  \plate {} { %
    (P1) %
   % (muj)(muj-f)(muj-f-caption) %
    (Sj)(Sj-f)(Sj-f-caption) %
  } {$j=1,\dots,J$} ; %
  
  % left-hand side titles  
  \node[text width=2.4cm, scale=1] at (-6.8,4.55) {second level};
  \node[text width=2cm, scale=1] at (-7.2,4) {hyperparameters};
  \node[text width=0.6cm, scale=1] at (-8.1,2.5) {hyperparameters};
  \node[text width=3cm, scale=1] at (-6,0.8) {parameters};
  \node[text width=0.15cm, scale=1] at (-7.4,-0.8) {observations};
% top title
% \node[text width=1.85cm, scale=1] at (-1.9,5.4) {DP($\alpha,G_0$)};   
\end{tikzpicture}
  \end{center}
\caption{A graphical representation of %MGMM 
MBA with conjugate priors. The overall mean, $\boldsymbol{\mu}$, and the precision matrix, $\Omega_e$, are assumed to be independent.} 
\label{fig:graphical_representationMGME}%
\end{figure}
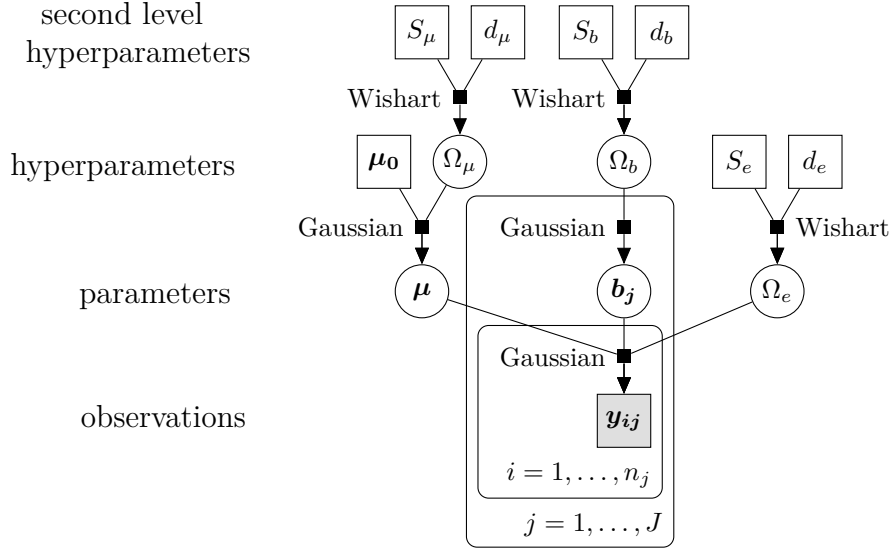

In \citet{sottas2007bayesian}, the T/E marker is modelled by a univariate Gaussian distribution in a Bayesian context. In this section, we present a multivariate Bayesian adaptive model (MBA), %\textbf{I'D SUGGEST MULTIVARIATE BAYESIAN ADAPTIVE MODEL INSTEAD OF MGMM}
which can deal with a wide variety of markers in their logarithmic scale as a generalisation of the univariate model of \cite{sottas2007bayesian}. The MBA is expressed as
\vspace{-0.2cm}
\begin{equation}
\label{equ:multi}
y_{ijk} = \mu_{k} + b_{jk} + e_{ijk}, \,\,\,\,  \boldsymbol{b_{j}} \overset{\text{iid}}{\sim} \mathcal{N}_{K}(\boldsymbol{0}, \Omega_{b}^{-1}),  \,\,\,\,  \boldsymbol{e}_{ij} \overset{\text{iid}}{\sim} \mathcal{N}_{K}(\boldsymbol{0}, \Omega_{e}^{-1}),
\end{equation}
\justify
where $y_{ijk}$ denotes the logarithm of the $i$th observation of the $k$th marker for the $j$th athlete, $\mu_{k}$ is the fixed effect for the overall mean of all observations of $k$th response marker, $b_{jk}$ is the random effect of athlete $j$ for the $k$th marker, and $e_{ijk}$ is the random term for other variation in its $i$th measurement, while $i=1,...,n_j$, $j=1,...,J$ and $k=1,...,K$. The assumptions here are that for a certain marker $k$, the random effects $b_{jk}$ are independent, identically and normally distributed between subjects and the error terms $e_{ijk}$ are also independent, identically and normally distributed between and within subjects. Shorthand notation for the overall mean is $\boldsymbol{\mu}=\{\mu_k\}_{k=1}^{K}$ and for the random effects is $\boldsymbol{b}= \{\boldsymbol{b_{j}}\}_{j=1}^{J} = \{b_{jk}\}_{j=1 \,\, k=1}^{J \,\,\,\, K}$. The precision matrices are denoted by $\Omega_{\mu}$ and $\Omega_{e}$ for the overall mean $\boldsymbol{\mu}$ and for the error term $\boldsymbol{e}_{ij}$, respectively. $\Omega_{b}$ is the precision matrix of  $\boldsymbol{b_{j}}$, which captures the correlation between the $K$ markers. % for each athlete.
Suppose we have the following Bayesian hierarchical multiple response model
\vspace{-0.2cm}
\begin{equation} 
\begin{split}
\boldsymbol{y}_{ij}|\boldsymbol{\mu_j} & \overset{\text{iid}}{\sim} \mathcal{N}_{K}(\boldsymbol{\mu_j}= \boldsymbol{\mu} + \boldsymbol{b_j} , \, \Omega_e^{-1})
\\
\boldsymbol{\mu}\, | \boldsymbol{\mu_0} & \overset{\text{iid}}{\sim} \mathcal{N}_{K}( \boldsymbol{\mu_0}, \, \Omega_{\mu}^{-1})  \\
\boldsymbol{b_j}  & \overset{\text{iid}}{\sim} \mathcal{N}_{K}( \boldsymbol{0}, \, \Omega_b^{-1}) \\
\Omega_e & \overset{\text{iid}}{\sim} \mathcal{W}i(d_e, S_e) \\
\Omega_{\mu} & \overset{\text{iid}}{\sim} \mathcal{W}i(d_{\mu}, S_{\mu})\\
\Omega_b & \overset{\text{iid}}{\sim} \mathcal{W}i(d_b, S_b),
\end{split}
\label{equ:model}
\end{equation}
\justify
where the overall mean $\boldsymbol{\mu}$ and the random effects $\boldsymbol{\mu_j}$ have conjugate multivariate Gaussian priors, while the precision matrices $\Omega_{e}$, $\Omega_{\mu}$ and $\Omega_{b}$ have conjugate Wishart priors. Historical prior information about all response variables is captured by the prior mean vector $\boldsymbol{\mu_0}$. Moreover, $d_{e}$, $d_{\mu}$ and $d_b$ denote the degrees of freedom, and $S_{e}$, $S_{\mu}$ and $S_b$ are the prior covariance matrices which are selected such that their corresponding prior distributions will be non-informative. The degrees of freedom of the Wishart distribution need to be greater than the data dimension minus one, i.e. $ d_{e} > K -1 $, so a non-informative value for this parameter is chosen to be equal to $K$ \cite{degroot2005optimal}. The prior covariance matrices $S_e$, $S_{\mu}$ and $S_b$ are all set equal to $(1/1000) \mathbb{I_K}$ so that posterior inferences would be largely driven by the data. A graphical representation of the MBA is shown in Figure \ref{fig:graphical_representationMGME}.

% Citation influence model
\vspace{-0.35cm}
\justify
We approximate the posterior parameters by using Gibbs sampling \cite{casella1992explaining, gelman2013bayesian}. To apply the Gibbs sampler (see Algorithm \ref{alg:Gibbs3}), the full conditional posterior distributions for each of the unknown parameters of $\boldsymbol{\theta} = (\boldsymbol{\mu},\{\boldsymbol{\mu_j}\}_{j=1}^{J}, \Omega_e,\Omega_{\mu}, \Omega_b)$ are calculated as follows: 
\begin{equation} 
\begin{split}
\boldsymbol{\mu} | \{\boldsymbol{\mu_j}\}_{j=1}^{J}, \Omega_e, \Omega_{\mu}, \Omega_b , \boldsymbol{y} & \sim \mathcal{N}_{K}(A_{n}^{-1} b_{n}, \, A_{n}^{-1}) 
\\
\boldsymbol{\mu_{j}} | \boldsymbol{\mu}, \boldsymbol{\mu_{-j}}, \Omega_e, \Omega_{\mu}, \Omega_b , \boldsymbol{y}  & \sim \mathcal{N}_{K}( A_{n}^{'-1} b_{n}', \, A_{n}^{'-1}) \,\,\,	\forall \,\, j = 1,...,J
\\
\Omega_e | \{\boldsymbol{\mu_j}\}_{j=1}^{J}, \boldsymbol{\mu}, \Omega_{\mu}, \Omega_b , \boldsymbol{y} & \sim \mathcal{W}i( d_e', S_e') 
\\
\Omega_{\mu} | \{\boldsymbol{\mu_j}\}_{j=1}^{J}, \boldsymbol{\mu}, \Omega_e,\Omega_{b} , \boldsymbol{y} & \sim \mathcal{W}i(d_{\mu}', S_{\mu}') \\
\Omega_b | \{\boldsymbol{\mu_j}\}_{j=1}^{J}, \boldsymbol{\mu},
\Omega_e,\Omega_{\mu} , \boldsymbol{y} & \sim \mathcal{W}i(d_b', S_b'),
\end{split}
\label{equ:posteriors}
\end{equation}
\justify
where in this section $\boldsymbol{y}$ is a $n_j \times J \times K$ data matrix and
\vspace{-0.3cm}
\begin{equation*} 
A_{n}= J\Omega_b + \Omega_{\mu},   \hspace{0.5cm}
b_{n} =\Omega_{\mu} \boldsymbol{\mu_0} + J\Omega_b\boldsymbol{\bar{\mu}}, \hspace{0.5cm}
\boldsymbol{\bar{\mu}} = \frac{1}{J}\sum_{j=1}^{J} \boldsymbol{\mu_j},
\end{equation*}

\vspace{-0.30cm}
\begin{equation*} 
A_{n}'=\Omega_b +n_j\Omega_e,   \hspace{0.5cm}
b_{n}' = \Omega_b\boldsymbol{\mu} + \Omega_e \sum_{i=1}^{n_j} \boldsymbol{y_{ij}},   
\end{equation*}

\vspace{-0.3cm}
\begin{equation*} 
d_e' =  d_e + n,   \hspace{0.5cm}
n = \sum_{j=1}^{J}n_j,   \hspace{0.5cm}
S_e' =  S_e^{-1} + \sum_{j=1}^J\sum_{i=1}^{n_j}(\boldsymbol{y_{ij}} -\boldsymbol{\mu_j})(\boldsymbol{y_{ij}}-\boldsymbol{\mu_j})^{T},   
\end{equation*}

\vspace{-0.3cm}
\begin{equation*} 
d_{\mu}' = d_{\mu} + 1, \hspace{0.5cm} S_{\mu}' =  S_{\mu}^{-1} + (\boldsymbol{\mu} - \boldsymbol{\mu_0})(\boldsymbol{\mu} - \boldsymbol{\mu_0})^{T},
\end{equation*}

\vspace{-0.3cm}
\begin{equation*} 
d_b' = d_b + J,  \hspace{0.5cm}  S_b' =  S_b^{-1} + \sum_{j=1}^J(\boldsymbol{\mu_j} - \boldsymbol{\mu})(\boldsymbol{\mu_j} - \boldsymbol{\mu})^{T}.   
\end{equation*}

\begin{algorithm}[H]
\caption{Gibbs algorithm}
\label{alg:Gibbs3}
\begin{algorithmic}[1]
\Require{ Generate an initial state $\boldsymbol{\theta}^{(0)} = (\boldsymbol{\mu}^{(0)},\{\boldsymbol{\mu_j}^{(0)}\}_{j=1}^{J}, \Omega^{(0)}_e, \Omega^{(0)}_{\mu}, \Omega^{(0)}_b)$} 
\For{$t \gets 1 \textrm{ to } T$}\\
draw $\boldsymbol{\mu}^{(t)} \,\, \sim p(\boldsymbol{\mu} \, | \{\boldsymbol{\mu}^{(t-1)}_j\}_{j=1}^{J}, \, \Omega_{e}^{(t-1)} , \Omega_{\mu}^{(t-1)}, \Omega_{b}^{(t-1)}, \boldsymbol{y})$
\For{$j \gets 1 \textrm{ to } J$}\\
\hspace{0.45cm}
draw $\boldsymbol{\mu_j}^{(t)} \,\, \sim p(\boldsymbol{\mu}_j \, | \boldsymbol{\mu}^{(t)}, \boldsymbol{\mu}_{-j}^{(t-1)}, \, \Omega_{e}^{(t-1)} , \Omega_{\mu}^{(t-1)}, \Omega_{b}^{(t-1)}, \boldsymbol{y})$ 
\EndFor \\
draw $\Omega^{(t)}_e \sim p(\Omega_e \, | \, \boldsymbol{\mu}^{(t)}, \{\boldsymbol{\mu_j}^{(t)}\}_{j=1}^{J}, \Omega_{\mu}^{(t-1)}, \Omega_{b}^{(t-1)}, \boldsymbol{y})$\\
draw $\Omega^{(t)}_{\mu} \sim p(\Omega_{\mu} \, | \, \boldsymbol{\mu}^{(t)}, \{\boldsymbol{\mu_j}^{(t)}\}_{j=1}^{J}, \Omega_{e}^{(t)}, \Omega_{b}^{(t-1)}, \boldsymbol{y})$ \\
draw $\Omega^{(t)}_b \sim p(\Omega_b \, | \, \boldsymbol{\mu}^{(t)}, \{\boldsymbol{\mu_j}^{(t)}\}_{j=1}^{J}, \Omega_{e}^{(t)}, \Omega_{\mu}^{(t)}, \boldsymbol{y})$
\EndFor
\end{algorithmic}
\end{algorithm}

\section{Methodology}
\label{OCC}

\subsection{One-class classification}
One-class classification (OCC) algorithms are used in classification when only one class (known as ``target'' class) is fully known and the others are either absent or poorly sampled \cite{minter1975single, bishop1994novelty, khan2014one}. Doping detection constitutes a challenging topic in forensic toxicology, and can be framed as a one-class classification problem since measurements from doped athletes can be difficult to obtain, either due to the elaborated techniques that athletes use to avoid testing, or due to the undetectable use of banned substances.  

\justify
For doping analysis, full information is provided on non-doped athletes who have been voluntarily tested, but limited knowledge is available for athletes who have received doping regimens. Thus, the samples from athletes with normal concentration values are treated as the ``target'' class. The focus is on studying whether there is evidence that new samples from athletes, whose doping status is unknown, are compatible with the known normal class of samples, or whether they show an abnormal pattern and should be considered as outliers. A classifier, that is a function which assigns each input data point to a class, accounting for other confounding factors such as sex, %gender \textbf{WOULD BE LESS AMBIGUOUS TO REFER TO SEX INSTEAD OF GENDER AS WE ARE DEALING WITH MALE/FEMALE}
cannot be constructed with known standard rules in the case of imbalanced classes. In a machine learning context, the main purpose is to infer a classifier from a limited set of training data noting that, in addition to the complexity of unbalanced data, the classifier should also have the capability to deal with longitudinal data, their updating nature as well as potential confounders. We approach the one-class classification problem using a density estimation method for OCC models in $K$-dimensional space as described in the following section. 

\subsection{Highest posterior predictive density}

The Bayesian model specification contributes to hierarchically shift from the prior evidence about population parameters $\boldsymbol{\theta}$ to the revised knowledge, expressed in the posterior density $p(\boldsymbol{\theta}|\boldsymbol{y})$, as new data become available. Using MCMC sampling methods, we can estimate the posterior density function and then approximate the $K$-variate predictive density function of a $1 \times K $ new observable vector $\boldsymbol{y_{n+1}}$ given the data $\boldsymbol{y}$ of dimensionality $n_{j'} \times J' \times K$, where $J'=J+1$. Therefore, the predictive density function is calculated as

\vspace{-0.1cm}
\begin{equation}
\label{predictive_density}
\begin{aligned}
p(\boldsymbol{y_{n+1}}|\boldsymbol{y=\{Y_0, Y\}})= \int_{\Theta} f(\boldsymbol{y_{n+1}} | \boldsymbol{\theta}) p(\boldsymbol{\theta}|\boldsymbol{Y_0, Y})d \boldsymbol{\theta},
\end{aligned}
\end{equation}
\justify
which is formed by weighting the possible values of $\boldsymbol{\theta}$ in the future observation $f(\boldsymbol{y_{n+1}} | \boldsymbol{\theta})$ by how likely we believe they are to occur, $p(\boldsymbol{\theta}|\boldsymbol{y})$. Note that $\boldsymbol{y}$ consists of both data from the baseline normal population, $\boldsymbol{Y_0}^{n_j \times J \times K}$, and all previous $n$ normal recordings of the under study athlete $J'$, $\boldsymbol{Y}^{n \times K}=(\boldsymbol{y_1},...,\boldsymbol{y_n})^\top $. We can use the predictive distribution to provide a useful range of plausible concentration values for a set of $K$ markers and ratios of a future athlete. Specifically, $\boldsymbol{y}$ is the training set and consists of the samples from the ``target'' class; that is the concentration values from non-EAAS users, which are considered to be within the normal range. To overcome the curse of dimensionality, we need to ensure a large number of observations in the training set. The main task of the OCC algorithm is to define a classification boundary, such that it accepts as many samples as possible from the normal class, while it minimises the chance of accepting the outlier samples. Hence, the classification is performed by setting a threshold value, $\gamma$, on the approximated densities, in such a way that a target (normal) and a non-target (outlier/abnormal) region can be obtained ensuring a low predefined Type I error (false positive rate) $\alpha$. Therefore, the $100(1-\alpha)\%$ prediction interval for $\boldsymbol{y_{n+1}}$ is the region of the form

\vspace{-0.1cm}
\begin{equation}
\label{HPP_region}
\begin{aligned}
C_{\alpha} = \{ \boldsymbol{y_{n+1}} \in \mathbb{R}^{K} : p\boldsymbol{(y_{n+1}}  | \boldsymbol{y}) \geq  \gamma \},
\end{aligned}
\end{equation}

\justify
where $\gamma$ is the largest constant such that Pr$(\boldsymbol{Y_{n+1}} \in C_{\alpha} | \boldsymbol{y}) \geq 1-\alpha$ \cite{hyndman1996computing}. %, as shown in Figure \ref{fig:HPDI_plot}.
%a predefined sensitivity; i.e. the proportion of identifying correctly the outlier samples. % ot target samples?
A new test result $\boldsymbol{y_{n+1}}$ is considered to be an outlier if, for at least one biomarker $k$, the corresponding observable $y_{n+1}^{(k)}$ is not included in the $100(1-\alpha)\%$ highest posterior density (HPD) intervals %percentile range in Sottas
of its conditional
probability distribution $p_{k}(y_{n+1}^{(k)} |\boldsymbol{y_{0}}^{(k)  }, %\, n_j \times J
y_1^{(k)},...,y_n^{(k)})$, and normal otherwise. %An indicator variable $\mathcal{X}$ is used to classify each data point $i$ as
%\caption{Highest posterior predictive density region of a unimodal posterior distribution $p(y_{n+1} | y_1,y_2,...,y_n)$. The grey shaded area denotes the region $C_{\alpha}$ where normal measurements from the athletes' steroid profile are expected to lie with probability $1-\alpha$. Observations which lie outside the $(1-\alpha)\%$ HPD interval are treated as outliers.}
%\vspace*{-0.1cm}
%\begin{equation}
%  \mathcal{X}_{i} =
%    \begin{cases}
%      1, & \text{if  } y_i \in C_{\alpha}   \\
%      0, & \text{otherwise.}
%    \end{cases}     
%\end{equation}
%\justify
Based on the %threshold, $\gamma$, 
decision rule in (\ref{HPP_region}), lower and upper limits of the HPD intervals are obtained, which define the normal boundaries of the EAAS concentrations or ratios at an individual level. These boundaries can be used in detecting any steroid misuse that may cause abnormally high or low concentration values of biomarkers or ratios, as well as in revealing urine samples replacement or the impact of other confounding factors. It is worth mentioning that there is the usual trade-off in choosing an appropriate $\alpha$, since lower $\alpha$ values give wider intervals. High $\alpha$ values give narrower intervals implying that an extreme new measurement % $\boldsymbol{y_{n+1}}$ 
has a low probability of lying in the interval. Furthermore, note that testing the first measurement of a new athlete is based only on the population thresholds, since $n=0$. Population thresholds are presented in Table \ref{table:WADAreflimits} and were obtained by \citet{van2010reference}. %\footnote{(This is not based solely on the data in citation 13, but on several studies on the distribution of endogenous steroids have been carried out. We may stop the sentence after Table 7 and use the citation 13 (and another one) to compare the data of our reference data and to show that we used normal values here?) } 
Further information about population thresholds was derived from the work of  \citet{rauth1994referenzbereiche} and \citet{kicman1995proposed}.
\justify
%We focus on the question of inferring the data to evaluate the hidden associations between them and the model parameters. 
%As for the univariate case, semi-supervised learning techniques are used, i.e. the one-class classification (OCC) method, to train the multivariate model in (\ref{equ:model}). The OCC method was applied to train the multivariate model in (\ref{equ:model}) using observations only from the normal class %(i.e. the majority class) as training set. The MGMM model is used to extract information from the high-dimensional probability distribution and classify the new ``unlabelled'' data from athletes to the most likely class, either to the normal class (\textit{non-suspicious}), if they adapt smoothly in the distribution of the ``normals'', or to the abnormal class (\textit{suspicious}).
%\\\\

\subsection{Continuity assumption}

In this semi-supervised learning process, we assume that the continuity assumption holds. This is a general assumption
in pattern recognition, according to which points that are close to each other are more likely to share a label. For this purpose, when the model suggests an outlier, then this observation is automatically excluded from the set of recordings that are used to compute the HPD intervals for normal samples. If we do not discard the observations flagged up as outliers, we should expect to learn the noise. Any noise measurements which are considered as normal measurements have a significant impact on the personalised accepted limits. Hence, we cannot expect to infer a good classification in such a case. 

\subsection{Dealing with imbalanced classes}

There is a variety of methods for imbalanced binary classification problems, the main goal of which is to convert the imbalanced dataset into balanced distributions by altering the size of the original dataset to provide the same proportion in each class. We chose to work with the method of \textit{``Random Oversampling''}, which randomly replicates the observations from the minority class to balance the data \cite{kotsiantis2006handling}. The ROSE package in R was used to generate replicates of the data from each athlete \cite{lunardon2014rose}. 

%\begin{figure}[H]%
%\centering
%\includegraphics[scale=0.55]{figures/HPDI.png}
%\caption{Highest posterior predictive density region of a unimodal posterior distribution $p(y_{n+1} | y_1,y_2,...,y_n)$. The grey shaded area denotes the region $C_{\alpha}$ where normal measurements from the athletes' steroid profile are expected to lie with probability $1-\alpha$. Observations which lie outside the $(1-\alpha)\%$ HPD interval are treated as outliers.}
%\label{fig:HPDI_plot}
%\end{figure}

\section{Results}

\subsection{Data summary}
\begin{figure}[H]%
\centering
%\subfigure {  % []
\includegraphics[scale=0.27]{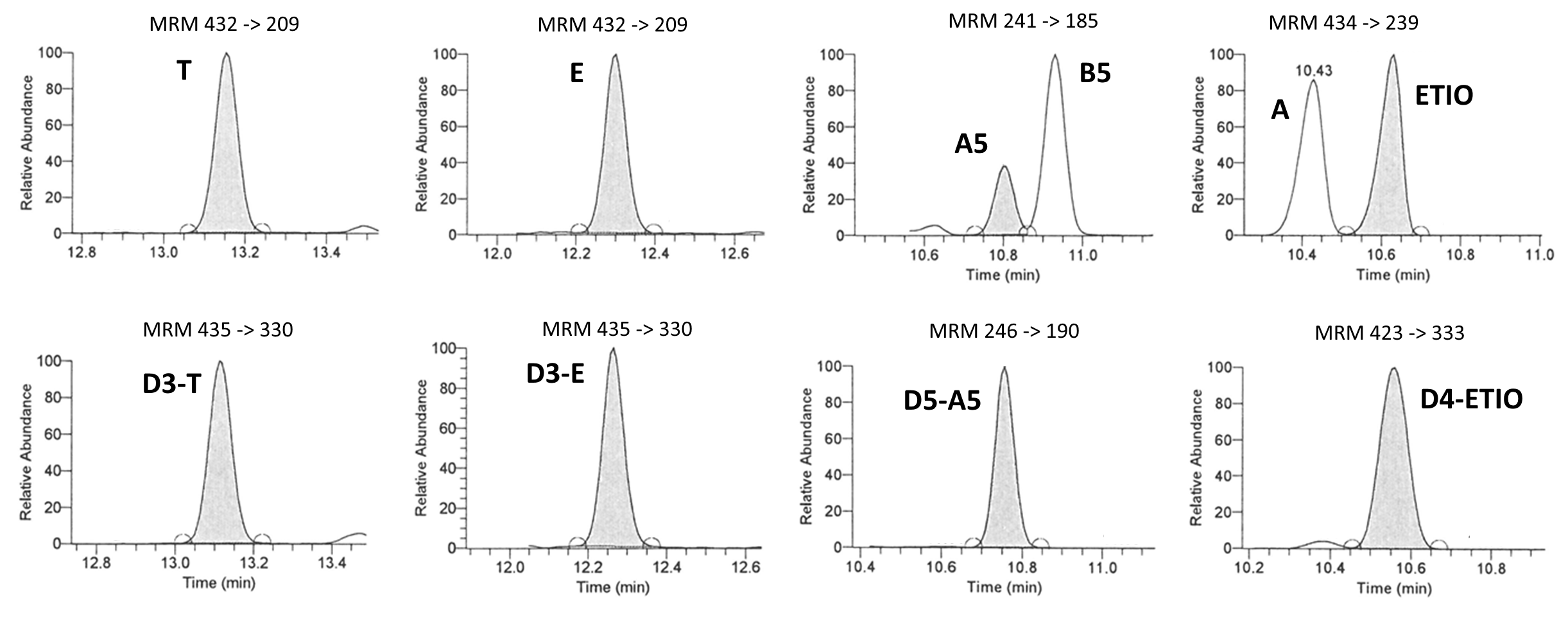}%}  
%\qquad
%\subfigure{   % []
%\includegraphics[scale=0.4]{figures/Capture2.png}
%}
\caption{Extracted chromatograms obtained for an unsuspicious urine sample. Shown are the multiple reaction monitoring (MRM) ion transitions employed for the quantification of endogenous steroids (upper part) and their deuterated analogues (lower part).}
\label{fig:chromatogr}
\end{figure}

The proposed method was applied to athletes' longitudinal steroid profile data extracted from their ABP. The datasets have been collected by following all the appropriate ethical approval procedures. Individual steroid profiles were analysed according to established methods including gas chromatography-mass spectrometry (GC-MS). Figure \ref{fig:chromatogr} represents a real GC-MS multiple reaction monitoring chromatogram produced by an unsuspicious urine sample. %\footnote{Any comments here?} %(I am not sure if it really makes sense to put here chromatograms, as these are not really meaningful.They look more or less the same - that´s why we need a better data evaluation. I may put together an exemplarily layout of the ion transitions we use?)} 
The longitudinal dataset includes six endogenous androgenic steroid concentrations and five concentration ratios proposed by WADA (i.e. T, E, A, Etio, A5, B5, T/E, A/T, A/Etio, A5/B5 and A5/E), which were repeatedly collected from each athlete in or out-of-competition. %In each of the two groups of normal and atypical samples, fifty male and fifty female athletes were negative, while 15 male and 14 female athletes were positive with at least one confirmed abnormal sample in their steroid profile, employing IRMS in line with WADA regulation \cite{WADATD2021IRMS}. %(pleaase cite here the TD IRMS: https://www.wada-ama.org/sites/default/files/resources/files/td2021irms_final_eng_v_2.0.pdf). 
%chromatogram plots
\justify
A total of 1433 normal urine samples were obtained from 100 athletes (50 males and 50 females), 2504 samples were obtained from 100 athletes (50 males and 50 females) whose longitudinal steroid profiles contain values classified as atypical, and 462 samples from 29 athletes (15 males and 14 females) with at least one confirmed abnormal value in their steroid profile, employing IRMS in line with WADA regulation \cite{WADATD2021IRMS}. Figure \ref{fig:ggpairsplot_normalvsabnormal} in Appendix \ref{figures} shows that there is %evidence of 
severe imbalance between normal (grey) and non-normal (black) samples with the former specifying the majority class. % (or the ``target'' class.), while the minority class (or ``non-target'' class) consists of atypical and abnormal samples. 
Specifically, out of 4399 urine samples across all athletes, only 327 (7.43\%) were non-normal values (275 from athletes with atypical samples, and 52 from athletes with abnormal samples). \begin{comment}
Since we deal with a severely skewed class distribution, we apply the majority impact learning technique, known as one-class classification, to all models presented in Sections \ref{univarmodel} and \ref{multivmixed} as discussed in Section \ref{OCC}. %, as well as an unstructured nature of the minority class. This indicates a significant imbalance between negatively and positively tested athletes %with normal/atypical and abnormal samples, respectively. %, as shown in Figure \ref{fig:barplot}.,Specifically, athletes belonging to the normal samples group were tested 6 to 47 times (14 times on average), athletes with atypical values were tested 6 to 69 times (25 times on average), and between 3 to 35 times (16 times on average) for athletes with abnormal samples. 
\end{comment}
Sample calibration was carried out prior to the analysis according to the estimated real limits of the applied methodology. Limit of detection (LOD) values and limit of quantification (LOQ) values within the steroid profiles of the athletes have been replaced by commonly accepted minimum cut-off values for all markers; i.e. all $<$LOQ and $<$LOD values in testosterone and epitestosterone were replaced by 1 ng/mL and 0.1 ng/mL respectively, while for $<$LOQ and $<$LOD values in the -diols were replaced by 5 ng/mL and 1 ng/mL, respectively.

\begin{comment}
\justify
Figures \ref{fig:series_markers} and \ref{fig:series_ratios} depict the variation of the values from the six biomarkers and their five ratios by gender against the sampling time (red: female, blue: male). Every trajectory represents measurements from a single athlete. The curves are separated into two classes; negative (non-doped) athletes, where no abnormal measurements are included in their steroid profiles, and positive (doped) athletes, whose steroid profile includes at least one abnormal measurement. Promising biomarkers and ratios can already be noticed from distinguishable trends in their trajectories. For example B5, T, T/E, A/T, and A5/E show a distinctive behaviour between the two classes. However, there are markers and ratios for which the difference between negative and positive athletes, either simply doesn't exist or is less obvious.
\end{comment}

\begin{figure}[H]%
\centering
\subfigure { %[]
\includegraphics[scale=0.5]{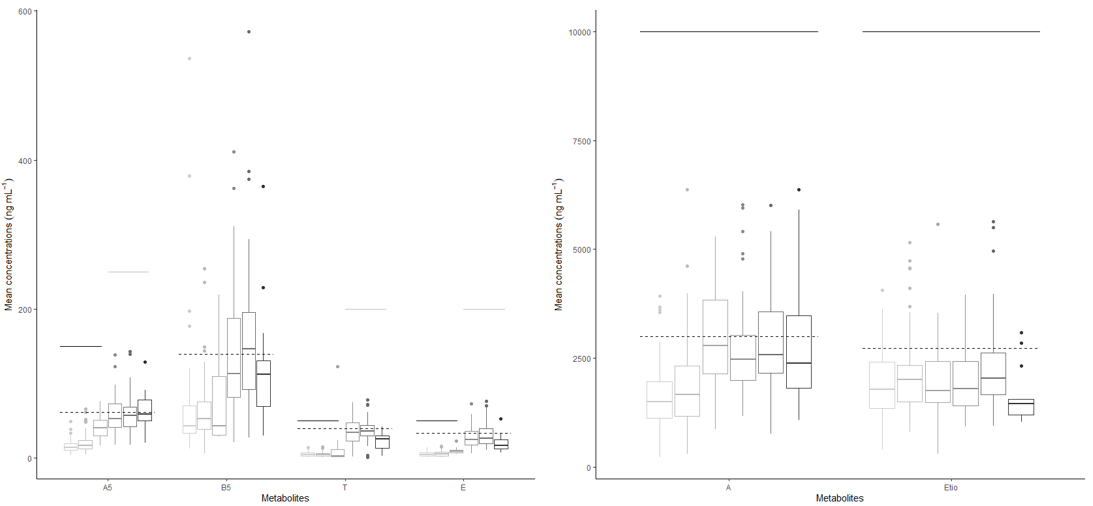}}  
\qquad
\subfigure { %[]
\includegraphics[scale=0.5]{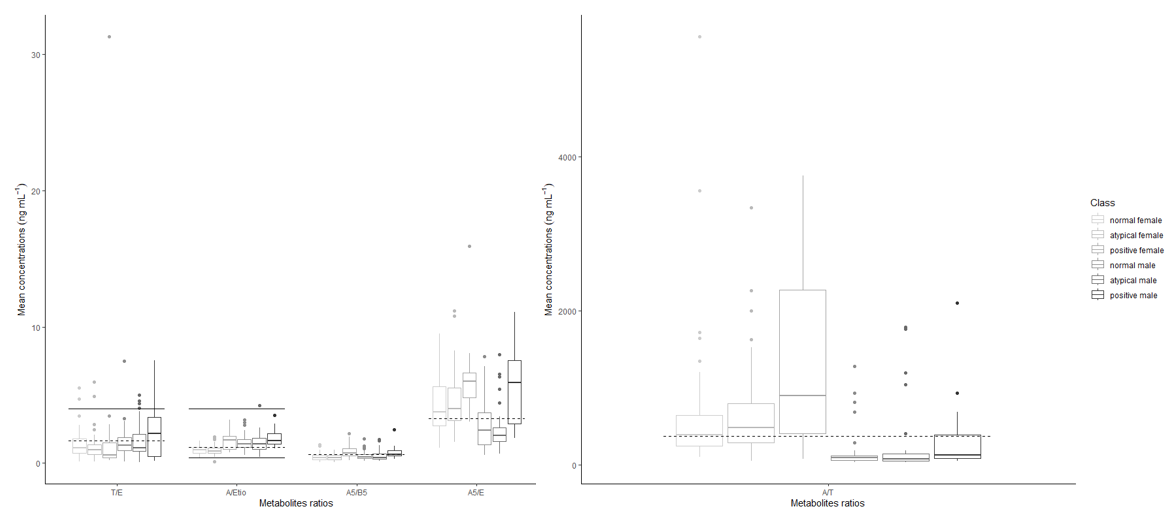}
}
\caption{Boxplots for the mean values of the metabolite concentrations (ng mL$^{-1}$) and their ratios from 229 athletes by sex %gender
and doping status (normal, atypical and abnormal). WADA limits are denoted by the solid lines. Dashed lines represent the mean concentrations of the metabolites obtained from the baseline population.}
\label{fig:mean_conc}
\end{figure}

\justify
Furthermore, single EAAS and ratio measurements of 164 healthy individuals have been provided as part of a cross-sectional study, representing a baseline population of which 91 were men and 73 women of age between 18 and 54 \cite{alladio2016application}. Figure \ref{fig:ggpairsplot} in Appendix \ref{figures} presents the scatter and density %and contour 
plots for men (black) and women (grey) for all available markers and ratios. The concentrations of markers seem to be slightly separable between men and women. However, this does not apply to the ratios, where the distributions of both sexes %genders \textbf{SEXES OR MALES AND FEMALES}
seem similar, except for the A/T ratio. The correlation between the various markers and ratios %is presented in Figure \ref{fig:correlograms}, which 
shows that plain markers are more highly correlated compared to the ratios (Figure \ref{fig:ggpairsplot}). 
\justify
Tables \ref{table:164healthysummary}, \ref{table:summary_long}, \ref{table:summary_long2} and \ref{table:summary_long3} summarise the statistics (minimum, %inter-quartile range; 
1st quartile; Q1 and 3rd quartile; Q3, %\textbf{SAY WHAT IQ1 AND IQ3 STANDS FOR}, 
mean, median, maximum EAAS values and standard deviation) of the baseline population and the longitudinally-monitored athletes. The boxplots of the mean values of the available longitudinal metabolites and ratios %obtained from both datasets with 4399 and 164 urine samples respectively, 
are reported in Figure \ref{fig:mean_conc}. When available, the WADA limits (lower and/or upper) are denoted by the solid lines based on Table \ref{table:WADAreflimits}. Androsterone and Etiocholanolone share the same population thresholds among females and males. When population-specific information is unavailable for certain ratios, the Q3 values derived from the cross-sectional dataset are used as initial population thresholds for the respective ratios. %Dashed lines represent the mean concentrations of the metabolites and ratios obtained from the baseline population.
%Gold diamonds signify the WADA threshold limits (lower and upper), which are available for some metabolites and ratios. 

\subsection{Estimation of highest posterior predictive density}

\subsubsection{Univariate HPD estimation}
To implement the univariate Bayesian model for each athlete and each biomarker in the ABP separately, we initially specified the prior distributions for the model parameters as described in Equations (\ref{priormean}) and (\ref{priorprecision}). While the correlation between the empirical mean and precision of log-transformed concentration values across all markers was relatively weak, ranging from -0.33 to 0.32, %indicating a weak association, the corresponding statistical tests performed on the 11 available markers and ratios support statistically significant correlation. Hence, 
we consider a prior dependency between the parameters $\mu$ and $\tau$. We set the hyperparameters $\kappa_0=1$, $\alpha_0=10$, and $\beta_0=1$, while we control for possible confounding due to sex %gender \textbf{SEX} 
differences by setting $\mu_0= \mu_{js}$, that is the mean of the $j$th marker in its logarithmic scale for male subjects if $s=0$, and for females if $s=1$, obtained from the baseline cross-sectional dataset.  %= \mu_y$. 
After the burn-in period of 1000 draws, 5000 draws for each parameter were sampled from the known joint posterior distribution (Gaussian-Gamma). The out-of-sample predictive distribution of a new test result $y_{n+1}$ given the previous recordings $\boldsymbol{y}=(y_1, y_2, ..., y_{n})$ is computed as 
%%heck rephrase?: 
%Note that we control for possible confounding due to gender differences by specifying the prior mean vector as the biomarkers' means of the baseline population, that is $\boldsymbol{\mu_0^{c}}=\boldsymbol{\bar{Y_0}^{c}}$, where $c=0$ for male and $c=1$ for female athletes. (figure 1?)
\begin{equation}
p(y_{n+1}|\boldsymbol{y}) = \int \int p(y_{n+1}|\mu,\tau) p(\mu,\tau|\boldsymbol{y}) d\mu d\tau \approx \frac{1}{T} \sum_{t=1}^{T} \bigg(\frac{\tau^{(t)}}{2\pi}\bigg)^{1/2} e^{- \frac{\tau^{(t)}}{2 } (y_{n+1} - \mu^{(t)})^2},
\label{eq:eq31}
\end{equation}

\justify
where $n \geq 0$, and $(\mu^{(t)}, \tau^{(t)})$ is the parameter pair of the $t$th draw obtained through the sampler with total number of iterations $T=5000$.  In practice, the integration averaging is performed using an empirical average based on samples from the posterior distribution. At first, we simulate replicates of new data, $y_{n+1}$, from the posterior predictive distribution and then we derive the $95\%$ HPD interval. % percentile range. 
%The model is applied on the 4399 EAAS concentrations and ratios of 229 athletes. 
Since the true class for each of the 4399 EAAS concentrations of the 229 athletes is known, we can estimate the predictive accuracy of the method. %Since the proportion of non-normal EAAS for the three groups of athletes is known (0/1433 for normal, 275/2504 for atypical, and 52/462 for abnormal) we can estimate the predictive accuracy of the method.

\begin{figure}[H]%
\centering
\subfigure[] {
\includegraphics[scale=0.32]{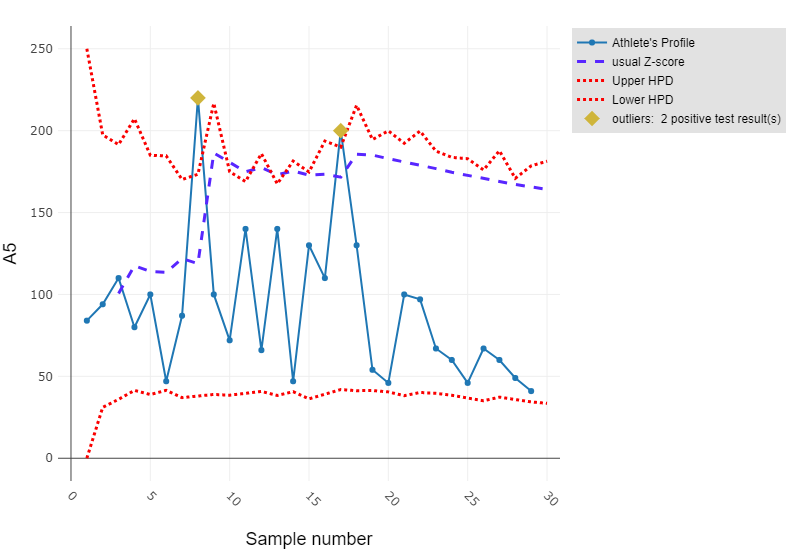}}
\subfigure[] {
\includegraphics[scale=0.32]{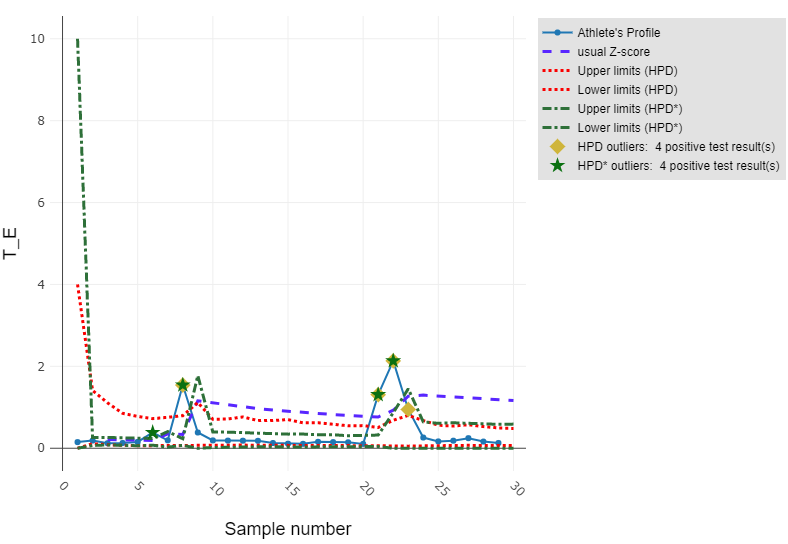}}
\vspace{-0.2cm}
\caption{A series of 29 longitudinal values of (a) A5 and (b) T/E (blue solid-dotted line) obtained from a doped athlete; the upper and lower limits (red dotted lines) are the 95\% HPD intervals from the univariate Bayesian model; upper limits based on a usual Z-score (purple dashed line); abnormal values are marked with gold diamonds. For (b), upper and lower limits (green dashed-dotted lines) are the 95\% HPD interval from the T/E model of \citet{sottas2007bayesian}, with its suggested abnormal values indicated by green stars.}
\label{fig:positiveprofiles_A5_TE} %male
\end{figure}

\justify
In Figure %Figures \ref{fig:negativeprofiles} and 
\ref{fig:positiveprofiles_A5_TE}, the A5 and T/E series of %a non-doped and 
a doped athlete are depicted %, respectively, 
with the blue-solid lines. The red dotted lines are the 95\% HPD intervals of the predictive distribution, which serve as the posterior normal boundaries at each specific time point. Before observing any data, the upper limits are defined by WADA's population thresholds, when they are available (see Table \ref{table:WADAreflimits}). For marker B5, we used the maximum value obtained by the Caucasian population in \citet{van2010reference}, while for the remaining ratios (A5/B5, A5/E and A/T) we chose the Q3 values 4, 10 and 10,000, respectively, as reasonable starting thresholds derived from the cross-sectional dataset. The purple dashed lines indicate the usual Z-score upper limits as presented in \citet{sottas2007bayesian}. The gold diamonds symbolise the abnormal values in the athlete's profile, which the model considers as suspicious values that need further investigation. For the T/E ratio in  Figure \ref{fig:positiveprofiles_A5_TE}(b), there are two additional green dashed-dotted lines that indicate the upper and lower limits of the T/E model with informative priors introduced by \citet{sottas2007bayesian}. For example, in Figure \ref{fig:positiveprofiles_A5_TE}(a), there are two androsterone samples which are higher than the upper limits, and in Figure \ref{fig:positiveprofiles_A5_TE}(b), Sottas' model identifies four abnormal T/E tests, out of which three are in common with those suggested by the general univariate model. %For example, in Figure \ref{fig:negativeprofiles}(e), there are two testosterone samples which are lower than the lower boundaries, and in Figure \ref{fig:negativeprofiles}(g), Sottas' model identifies six abnormal T/E tests, while the general univariate model suggests five. 
%In general, considering that Athlete 7 has not received any doping substances, there were many false positives leading to a weak classification performance. 
%Regarding Athlete 202, whose 21$^{st}$, 22$^{nd}$ and 23$^{rd}$ sample tests are confirmed as abnormal, only E, T/E and A5/E were sensitive enough to detect these anomalies within the athlete's steroid profile. 

 \vspace{-0.25cm}

 \justify
 Figure \ref{fig:positiveprofiles} (a-k) displays the series of the six EAAS along with their five corresponding ratios for the same athlete. Given that the 21$^{st}$, 22$^{nd}$ and 23$^{rd}$ sample tests of that athlete are confirmed as abnormal, only E, T/E and A5/E were sensitive enough to detect these anomalies within the athlete's steroid profile. 
%\justify
Note that if the model suggests a sample as an outlier, we automatically exclude it from the set of recordings which are used to compute the HPD intervals, because it might have an impact on the validity of the following personalised normal limits. 

%\justify
\subsubsection{Multivariate HPD estimation}
\justify
To apply the multivariate Bayesian adaptive model, we initially specified the prior distributions for the model parameters $\boldsymbol{\theta}= (\boldsymbol{\mu},\{\boldsymbol{\mu_j}\}_{j=1}^{J}, \Omega_e, \Omega_{\mu}, \Omega_b)$ as described in Section \ref{multivmixed}. The prior covariance matrices $S_e$, $S_{\mu}$ and $S_b$ are all set equal to $(1/1000) \mathbb{I_K}$, and the degrees of freedom are $d_e=d_{\mu}=d_b=K-1$, where $K$ is the dimensionality of the data. We use historical prior information obtained from the baseline cross-sectional dataset of 164 non-doped athletes (91 men and 73 women), which is captured by the prior mean vector $\boldsymbol{\mu_{0}}$. Similarly, as in the univariate case, the model can accommodate distinct prior mean vectors for both men and women.
%Note that again as in the univariate case the model accommodates different prior mean vectors for men and women. 
Then, $3000$ draws were sampled for each parameter from the posterior distribution $p(\boldsymbol{\theta}|\boldsymbol{y})$ using the Gibbs sampler, while the first $1/3$ was discarded. Given the remaining set of samples $\{\boldsymbol{\theta^{(t)}}\}_{t=1}^{T=2000}$, our estimate for the predictive distribution is
\vspace{-0.15cm}
\begin{equation}
p(\boldsymbol{y_{n+1}}|\boldsymbol{y}) \approx \frac{1}{T} \sum_{t=1}^{T} %p(y_{n+1} | \boldsymbol{\theta^{(t)}}).
%f(\mathbf{x}; \boldsymbol{\mu_j}, \boldsymbol{\Omega}) = 
(2\pi)^{-K/2}|\Omega_e^{(t)}|^{-1/2}\exp\left(-\frac{1}{2}(\boldsymbol{y_{n+1}} - \boldsymbol{\mu_j^{(t)}})^T \Omega_e^{(t)} (\boldsymbol{y_{n+1}} - \boldsymbol{\mu_j^{(t)}})\right)
\label{eq:pred_distr_mult}
\end{equation}

\justify
We first simulate from the posterior predictive distribution many replicates of the new data, $\boldsymbol{y_{n+1}}$, and thus we derive the 95\% HPD interval. %The model is applied on the 4399 EAAS and ratios of 229 athletes (100 athletes with normal samples, 100 athletes with atypical samples and 29 athletes with abnormal samples). 
Data from athletes with exclusively normal samples were used for model training, whereas atypical and abnormal samples comprised the test set. The idea is to train the model with normal data from non-doped athletes, by estimating $p(\boldsymbol{\theta}_{\text{normal}}|\boldsymbol{y_{\text{normal}}} )$ and then test how likely it is for a future unlabelled observation to be generated by this model. %Table \ref{table:classif1} includes the classification performance of the proposed multivariate model applied to: a) all EAAS markers and ratios; b) EAAS markers only; and c) ratios only.

\subsection{Classification performance}

%In this section we present the classification performance of the developed models applied on the same dataset for detecting abnormal cases within athletes' steroid profiles. 
In forensic toxicology, high specificity is important, thus a very low false positive rate is required in order to prevent the accusation of an innocent athlete. However, classification accuracy values and measures regarding the majority class such as the overall accuracy, tend to be pretty high because they are computed under the assumption of balanced class distributions. Consequently, we need to use appropriate metrics for evaluating the classification performance of the models which can deal with the imbalance of the dataset, such as the F1 score and G-mean (Geometric mean) defined as 
\begin{equation}
%F_{\beta} = \frac{{(1 + \beta^2) \cdot \text{Precision} \cdot \text{Recall}}}{{\beta^2 \cdot \text{Precision} + \text{Recall}}}, 
F1\text{-score} = \frac{2 \times \text{Precision}  \times \text{Recall}}{\text{Precision} + \text{Recall}},
\end{equation}

\justify
and 

\begin{equation}
G\text{-mean} = \sqrt{\text{Sensitivity} \times \text{Specificity}}
\label{gmean}
\end{equation}

%\textbf{DEFINED AS ...}.

\justify
Table \ref{table:classif1} presents the classification performance of the univariate model and the MBA model %MGMM 
(before and after oversampling), in comparison with the univariate T/E model proposed by \citet{sottas2007bayesian}, the univariate Euclidean distance (ED) score model of \citet{de2023new} and a generalised linear mixed-effects model (GLMM) using a false positive rate $\alpha=0.05$. The GLMM model is implemented via the \texttt{glmer()} function in R to fit a mixed effects logistic regression model with a binomial distribution for the response variable ``Class$_{ij}$'', which is a binary response variable that represents the class of each observation $i$ of athlete $j$ including the biomarkers and/or ratios as individual level continuous predictors, and a random intercept by athlete ID. 
%%\textbf{WE DON'T DESCRIBE THE GLMM NOR DO WE MOTIVATE THE REASON IT IS INCLUDED IN THE RESULTS TABLE. COULD REMOVE AND REPLACE WITH UNIVARIATE MODEL FOR SCORE/EUCLIDEAN DISTANCE OR IF WE ARE KEEPING IT WE NEED TO EXPLAIN WHAT IT IS, HOW IT IS IMPLEMENTED AND WHY WE ARE USING IT WITHOUT OVERSAMPLING OR WITH SOME SORT OF REGULARISATION} \textbf{ADD UNIVARIATE MODEL WITH EUCLIDEAN DISTANCE, CITE ACA2023 PAPER FOR DETAILS} 
The classification using the univariate models %with false positive rate $\alpha=0.05$ 
suggest the T/E, A/ETIO and A5/E ratios as the most sensitive variables for detecting anomalies in the steroidal profile with higher $F1$ scores ($F1_{ {T/E}^1}= 0.19$, $F1_{A/ETIO}= 0.24$ and $F1_{A5/E}= 0.23$). Regarding the T/E models, the one employing informative priors demonstrates slightly better predictive performance in contrast to the model using semi-informative conjugate priors, as indicated by the improved metrics $F1_{ {T/E}^2}=0.20$, sensitivity$_{ {T/E}^2}=0.32$ and balanced accuracy$_{ {T/E}^2}=0.59$.
%The results from the T/E model in the first part of the table are based on semi-informative conjugate priors, whereas the results from the T/E model in the second part are based on the informative priors of \citet{sottas2007bayesian}. The latter model achieves a slightly better prediction performance based on the F1$_{ {T/E}^2}=0.20$, sensitivity$_{ {T/E}^2}=0.32$ and balanced accuracy$_{ {T/E}^2}=0.59$.

\begin{figure}[H]%
\centering
%\subfigure[] {  
\includegraphics[scale=0.48]{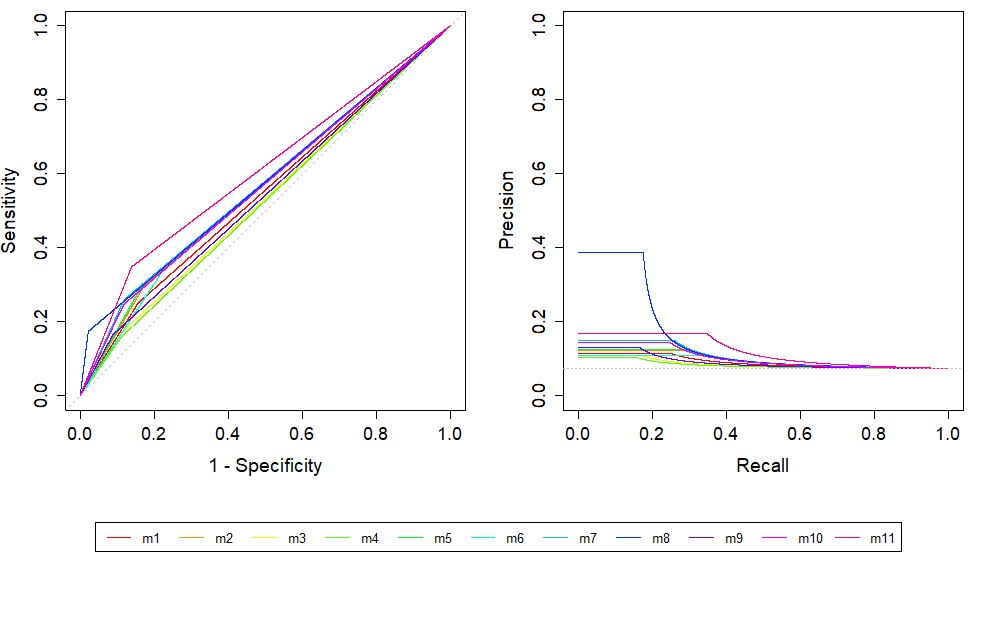}
\vspace{-1.4cm}
\begin{center}
    (a) 
\end{center}
%}
%\vspace{-0.5cm}
%\qquad
%\subfigure[]{  
\includegraphics[scale=0.51]{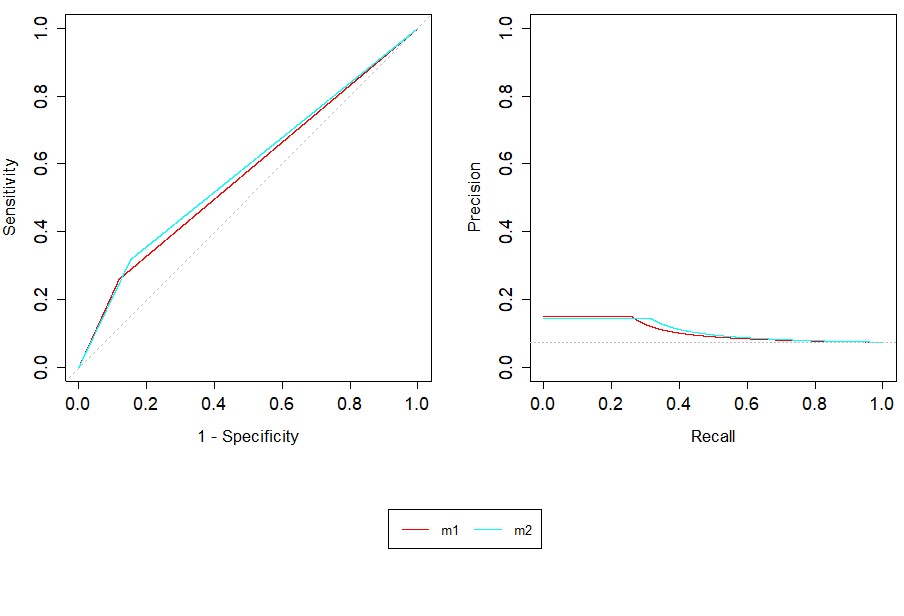}
\vspace{-1.3cm}
\begin{center}
    (b) 
\end{center}
%}
%\qquad
%\vspace{-0.5cm}
%\end{figure}
%\begin{figure}[H]%
%\centering
%\subfigure[]{  
\includegraphics[scale=0.46]{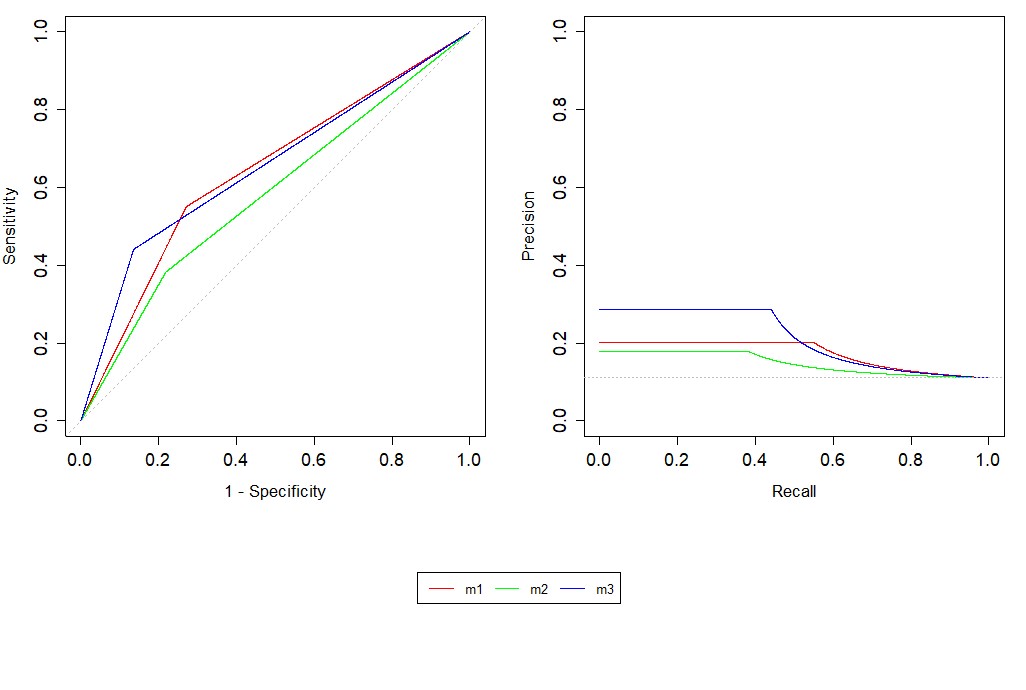}
\vspace{-1.4cm}
\begin{center}
    (c) 
\end{center}
%}
\vspace{-0.6cm}
\caption{ROC curves obtained from (a) the univariate models (m1: A5, m2: B5, m3: A, m4: ETIO, m5: T, m6: E, m7: T/E, m8: A/ETIO, m9: A/T, m10: A5/B5, m11: A5/E), (b) m1: T/E model vs m2: Sottas' model for T/E, and (c) the multivariate models (m1: all EAAS markers and ratios, m2: EAAS only, m3: ratios only) applied to the original dataset.}
\label{fig:ROC_plots}
\end{figure}

\justify
In Figure \ref{fig:ROC_plots}, we have also presented the ROC (Receiver Operating Characteristic) curves and the Precision-Recall curves to measure the accuracy of the classification predictions in the various models. According to the ROC curves, the model for the A5/E ratio showed superiority compared to the other univariate models in Figure \ref{fig:ROC_plots}(a), while the model for A/ETIO ratio showed superiority in the Precision-Recall plot. However, the superior performance of the A/ETIO ratio was unexpected based on the current literature, which is an aspect that warrants further investigaton. The curve using the model of \citet{sottas2007bayesian} for T/E is higher in Figure \ref{fig:ROC_plots}(b), which verifies its higher predictive performance.

\begin{table}[H] 
\centering 
\scalebox{0.82}{%
\begin{threeparttable}
\begin{tabular}{lllllllll}
\\ \hline 
\hline 
Classification & & &  & & & & Balanced & Overall \\ 
 model & Variable & $G$-mean & $F_{1}$ &  Precision  & Sensitivity & Specificity & Accuracy &  Accuracy (95\% CI)  \\ 
\hline 
Univariate & A5 & 0.46 & 0.15 & 0.11 & 0.25 & 0.84 & 0.55 & 0.80 (0.79, 0.81)\\ 
           & B5 & 0.50 & 0.17 & 0.12 & 0.30 & 0.83 & 0.56 & 0.79 (0.78, 0.80)\\ 
           & A  & 0.38 & 0.13 & 0.11 & 0.17 & 0.86 & 0.53 & 0.83 (0.82, 0.84)\\ 
           & ETIO & 0.38 & 0.12 & 0.10 & 0.16 & 0.89 & 0.52 & 0.84 (0.82, 0.85)\\
           & T  & 0.50 & 0.18 & 0.13 &0.30& 0.83 & 0.57 & 0.79 (0.78, 0.81)\\ 
           & E  & 0.52 & 0.17 &0.11 &0.34& 0.78 & 0.56 & 0.75 (0.73, 0.76)\\ 
           & T/E\tnote{1}  & 0.48 & 0.19 & 0.15 &0.26& 0.88 & 0.57 & 0.84 (0.82, 0.85)\\ 
           & A/ETIO & 0.41 & 0.24 & 0.39 &0.17& 0.98 & 0.58 & 0.92 (0.91, 0.93)\\ 
           & A/T  & 0.39 & 0.15& 0.13 &0.17 & 0.91 & 0.54 & 0.86 (0.85, 0.87)\\ 
           & A5/B5 & 0.47 & 0.18& 0.14 &0.25& 0.88 & 0.57 & 0.84 (0.82, 0.85)\\ 
           & A5/E  & 0.55 & 0.23 & 0.17 &0.35& 0.86 & 0.61 & 0.82 (0.81, 0.83)\\ 
\hline 
\citet{sottas2007bayesian} & T/E\tnote{2} & 0.52 & 0.20 & 0.15 & 0.32 & 0.85 & 0.59 & 0.81 (0.80, 0.82)   \\
\hline
pre-oversampling  &  EAAS & 0.55 & 0.24  & 0.18 & 0.38 & 0.78 & 0.58  & 0.74 (0.72, 0.75) \\ 
MBA   & ratios & 0.62  & 0.35  & 0.29  & 0.44 & 0.87 & 0.65 & 0.82 (0.80, 0.83)  \\ 
  &  all & 0.63 & 0.30  & 0.20  & 0.55 & 0.73  &  0.64 & 0.71 (0.70, 0.73)  \\ 
\hline   
post-oversampling    &  EAAS & 0.49 &  0.44 & 0.50  & 0.39 & 0.61 & 0.50 & 0.50 (0.49, 0.51)  \\ 
MBA & ratios & 0.46 & 0.38 & 0.50  & 0.31 & 0.69 & 0.50 & 0.50 (0.48, 0.52)   \\ 
               &  all & 0.50 & 0.49  & 0.50 & 0.47 & 0.53 & 0.50 & 0.50 (0.49, 0.52)   \\ 
\hline 
pre-oversampling  & EAAS & 0.22 & 0.20 & 0.11 & 0.98  & 0.05 & 0.52 & 0.15 (0.14, 0.17) \\
univariate    & ratios & 0.22 & 0.20  & 0.11  & 0.99  &  0.05 & 0.52 & 0.15 (0.14, 0.16) \\ 
ED score   & all & 0.20 & 0.20 & 0.11 & 0.99  & 0.04  & 0.52 & 0.15 (0.14, 0.16) \\ 
             \hline 
post-oversampling  & EAAS & 0.17 & 0.66 & 0.50 & 0.97  & 0.03 & 0.50 & 0.50 (0.49, 0.51) \\
univariate & ratios & 0.14 & 0.66  &  0.50  &  0.98 &  0.02 & 0.50 & 0.50 (0.49, 0.51) \\ 
ED score & all & 0.14 & 0.66 & 0.50 & 0.98  & 0.02  & 0.50 & 0.50 (0.49, 0.51) \\ 
               \hline               
pre-oversampling  & EAAS & 0.10 & 0.02 & 1 & 0.01  & 1 & 0.51 & 0.88 (0.87, 0.90) \\
GLMM   & ratios & 0.25 & 0.11 & 0.82 & 0.06  &  1 & 0.53 & 0.89 (0.87, 0.90) \\ 
%(pr>=0.80)  
& all & 0.25  & 0.12 & 0.83 & 0.06 & 1 & 0.53 & 0.89 (0.87, 0.90)\\ 
   \hline             
post-oversampling  & EAAS & 0.20 & 0.08 & 0.79 & 0.04  & 0.99 & 0.51 & 0.51 (0.49, 0.53) \\
GLMM    & ratios & 0.38 &  0.26  & 0.89   & 0.15  &  0.98 & 0.57 & 0.57 (0.55, 0.59) \\ 
%(pr>=0.80)  
& all & 0.39 &  0.27 & 0.81 & 0.16  & 0.96  & 0.56 & 0.56 (0.54, 0.58)\\ 
\hline   
\end{tabular} 
  \begin{tablenotes}
    \item[1] This model specifies weakly informative priors.
    \item[2] For Sottas' model the priors are set to be strongly informative.
  \end{tablenotes}
    \end{threeparttable}
  }
  \caption{\label{table:classif1} Predictive performance of the univariate and multivariate models on the athletes' profiles based on the 95\% HPD interval. }
\end{table} 

%\textbf{TABLE 1: FOOTNOTE NUMBERING FOR SOTTAS MODEL LOOKS LIKE T/E^2}

\justify
%multiv
The proposed multivariate Bayesian adaptive model (MBA) has been applied to: a) EAAS markers only; b) ratios only; and c) all EAAS markers and ratios; using both the original dataset and the dataset after over-sampling. The $G$-mean and $F1$ metrics for the MBA models, as shown in Table \ref{table:classif1} before oversampling, exhibit consistently higher values when compared to their counterparts in the univariate models. Comparing further between the performance of the MBAs, the five available ratios were found to be the most powerful set of variables before oversampling with the highest metric values $F1_{MBA_{ratios}}=0.35$, precision$_{MBA_{ratios}}=0.29$, sensitivity$_{MBA_{ratios}}=0.44$, specificity$_{MBA_{ratios}}=0.87$ and highest balanced accuracy$_{MBA_{ratios}}=0.65$. While the $G$-mean value for the model applied to the five ratios ($G\text{-mean}_{MBA_{ratios}}=0.62$), is slightly lower that of the model applied to the six markers and five ratios ($G\text{-mean}_{MBA_{all}}=0.63$), it is important to note that the model focusing on the ratios remains equally powerful despite utilizing less information. Similar conclusions about the superiority of the pre-oversampling multivariate model, assessed through the use of ratios, can be derived from an examination of the plots in Figure \ref{fig:ROC_plots}(c). Here, the blue lines point to a better relationship between sensitivity and 1-specificity, as well as between precision and recall.
%The same conclusions regarding the superiority of the pre-oversampling multivariate model using the ratios can be drawn by inspecting the diagrams in Figure \ref{fig:ROC_plots}, where the blue lines show a better relationship between sensitivity and 1-specificity, and between precision and recall. 

\begin{figure}[H]%
\centering
\subfigure {   %[]
\includegraphics[scale=0.45]{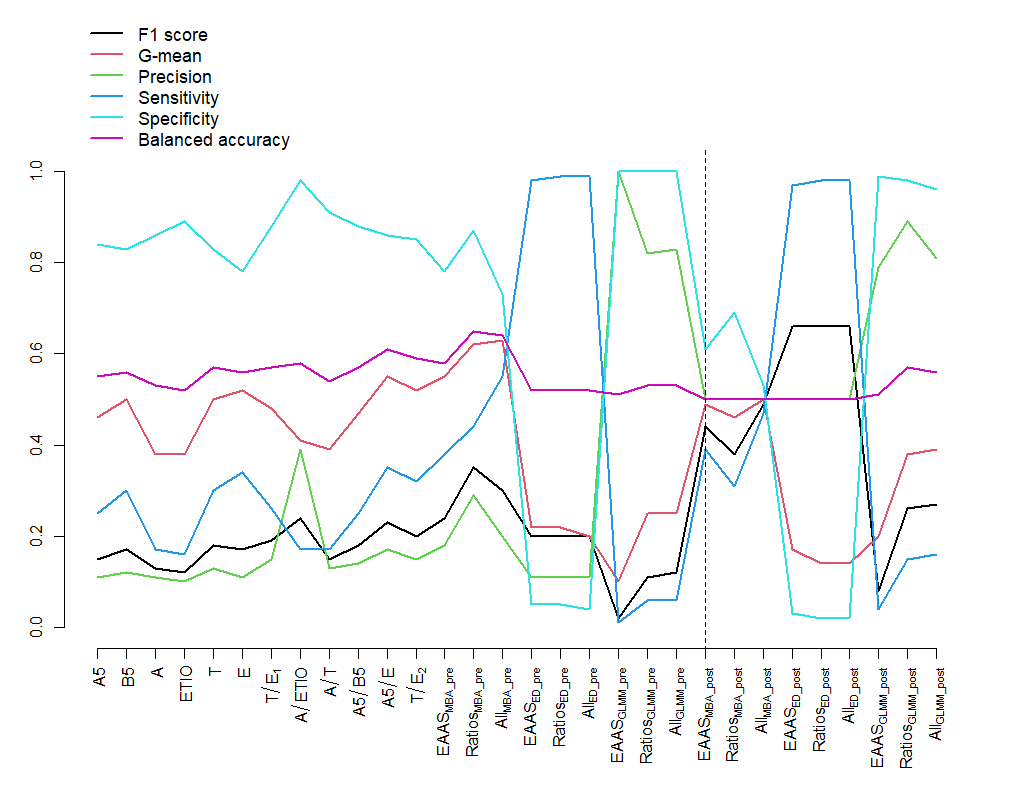}}  
%\qquad
%\subfigure{   % []
%\includegraphics[scale=0.30]{Figures/ROC_TE.png}
%}
\caption{Classification metrics per component; i.e. markers and ratios separately, only EAAS, only ratios, all EAAS markers and ratios. T/E$_1$ corresponds to the T/E model from Section \ref{univarmodel}, while T/E$_2$ corresponds to the T/E model by \citet{sottas2007bayesian}. Subscripts ``pre'', ``post'', denote pre-oversampling and post-oversampling, respectively.
}
\label{fig:Final_plot1}
\end{figure}

\justify
In Figure \ref{fig:Final_plot1}, all classification metrics are plotted for the applied models. %various combinations, i.e. each marker and ratio separately, markers only (EAAS), ratios only, and all markers and ratios together. 
Once more, the conclusion is clear that the multivariate Bayesian adaptive model (MBA) before oversampling applied solely to the ratios, consistently delivers the best overall classification performance ($F1_{MBA_{ratios}}=0.25$ and $G$-mean$_{MBA_{ratios}}=0.62$) compared to the univariate models, the univariate ED score models and the GLMMs.
%Again, we conclude that an overall best classification performance is achieved by applying the multivariate model (MGMM) on the ratios only without oversampling. 
The problem of unreliable overall accuracy in imbalanced data has been alleviated through the application of oversampling techniques, as evidenced by the now equivalent values of balanced accuracy and overall accuracy (see Table \ref{table:classif1}). Comparing the $F1$ scores before and after oversampling, it appears that applying the random oversampling method enhances the overall classification performance of the models. It is noteworthy that the univariate ED score model exhibits considerably higher sensitivity and lower specificity values, indicating a tendency to classify normal values as abnormal. Conversely, the GLMM performance is characterised by significantly higher specificity and lower sensitivity values, suggesting a propensity to classify abnormal values as normal. Therefore, their $F1$ score values are biased towards either very high or very low sensitivity. To address this, we employ the $G$-mean metric to compare all models after oversampling, which is the harmonic mean of sensitivity and specificity as defined in \ref{gmean}. After oversampling, the highest $G$-mean values are consistently achieved by the multivariate Bayesian adaptive model, which outperforms the univariate ED score model and GLMM.
%It is noteworthy that the performance of various GLMM applications was exceptionally poor. This can be attributed to the inherent instability of GLMMs when dealing with highly imbalanced sample sizes across different groups. %This happens because GLMMs can be unstable when sample sizes across groups are highly imbalanced. 

\vspace{-0.2cm}

\section{Discussion}

Our primary objective in this research was to develop a multivariate Bayesian adaptive model for repeated measurements of various urinary biomarkers and their ratios, as a generalisation of a widely used univariate model \cite{sottas2007bayesian} which also uses urinary steroid profile data for doping detection. Similar to the univariate model, the proposed methodology considers the population distribution of these biomarkers and the individual's own history; i.e. previous measurements of the athlete's biomarkers. In addition, it has the capacity to incorporate other relevant demographic characteristics such as the athlete's sex or age. The method simultaneously models multiple measurements of endogenous substances that are able to reveal the presence of doping agents, and utilises a one-class classification rule to improve detection performance in the presence of class imbalance. The resulting personalised normal ranges for athletes' longitudinal steroid profiles show improved detection performance in a dataset of professional athletes, as compared to the performance of the existing univariate model. The proposed method thus appears promising as an additional monitoring tool in the anti-doping toolkit. One of the method's advantages is that it relies on the urinary steroid profile, which produces easily accessible, reliable, quickly quantifiable and reproducible data from samples that are obtained in a non-invasive way and with a low financial cost. In addition, an implementation of this model is available as a user-friendly software designed specifically for anti-doping laboratories. 

\begin{comment}
\begin{figure}[H]
\begin{center}
\begin{tikzpicture}
\draw [thick, <->] (0,8) -- (0,0) -- (8,0);
\node [rotate=90] at (-1,4) {Sample size};
\node at (5,-1) {difference};
\draw [-stealth](3,-1.8) -- (6,-1.8);

\node [draw,rotate=90] at (1.5,4){standard classification / no OCC};
\node [draw, rotate=90] at (3,4) {simple OCC};
\node [draw] at (6,5) {OCC; subsampling};
\node [draw] at (6,3) {OCC; oversampling};
\end{tikzpicture}
\end{center}
\caption{Different classification methods for different sample sizes and for low, moderate and high values of difference in distribution between majority and minority class.} 
\label{fig:Proposal}%
\end{figure}
\end{comment}

\justify
%It is also worth mentioning that the proposed screening method is non-invasive, easily accessible, reliable, quickly quantifiable and reproducible. It also has low financial burden, low risk level, and achieves an improved predictive performance regardless of the imbalance of the data. These characteristics contribute effectively to improve doping detection in sports drug testing laboratories. 

A challenging aspect of this work was the presence of class imbalance due to the scarcity of data from confirmed doping cases. Oversampling was utilised to alleviate this problem, although other techniques such as undersampling, a combination of undersampling and oversampling or boosting algorithms that are able to convert weak learners to strong learners \cite{kotsiantis2006handling, schapire2013explaining} could also be explored. Applications on additional markers and ratios is another direction which is worth investigating. An illustrative example of this concept  can be found in the work of \citet{wilkes2018using}. %In Figure \ref{fig:Proposal}, a very coarse overview is given which shows a proposal of the applicability of different classification approaches to different situations. For small differences in distribution between majority and minority class, standard classification techniques work well for most of the cases regardless the sample size. For adequate sample sizes and moderate distribution differences simple, one-class classification can work well with respect to the predictive performance. For larger distribution differences, random oversampling in the OCC framework may work better when the sample size is small compare to the random subsampling, which can work better for larger sample sizes. \\\\
Lastly, this research examines the predictive performance of the Bayesian adaptive model when applied to six biomarkers and five ratios. It would be of interest to explore whether a subset of these variables could further improve performance. Further testing on additional datasets is needed to produce  robust recommendations as to the optimal implementation of the model. %Finally, to facilitate the practical use of the proposed method, we have integrated the necessary tools into user-friendly software designed specifically for anti-doping laboratories. %The tools for applying the proposed model are incorporated in a user-friendly software we have designed for use by anti-doping laboratories. 
%Including the age of athletes as a covariate in the model might enhance the model performance since the biosynthesis of endogenous androgenic anabolic steroids (EAAS) varies with age.

\section*{Acknowledgements}
%We are grateful to the Institute of Biochemistry of the German Sport University Cologne, which is an accredited laboratory of Anti-Doping Administration and Management System (ADAMS) of WADA, for sharing their laboratory datasets to carry out this research work. The data have been collected following all the appropriate ethical approval procedures. 
We are grateful to the EPSRC for supporting the PhD research of Dimitra Eleftheriou at the University of Glasgow. We would also like to thank Dr Ludger Evers for helpful discussions throughout this research work. The authors declare that there are no conflicts of interest.

\medskip
\bibliographystyle{unsrtnat} %\bibliographystyle{unsrt}
{\small \bibliography{Bibliography} } 
% The references (bibliography) information are stored in the file named "Bibliography.bib"

\newpage
\appendix
\section{Figures}
\label{figures}

\setcounter{figure}{0}
\renewcommand{\thefigure}{A.\arabic{figure}}

\vspace{-0.7cm}
\begin{figure}[!htbp]%
\centering
\subfigure[] {
\includegraphics[scale=0.46]{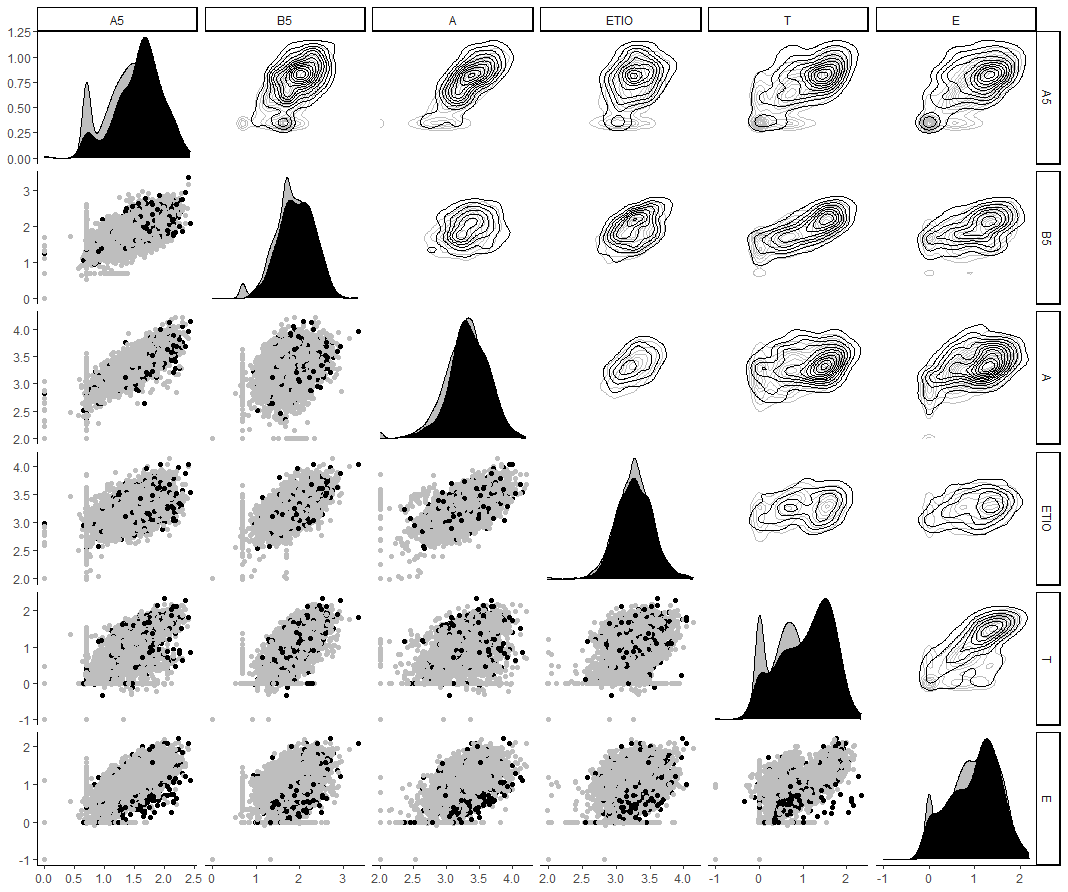}}
\qquad
\subfigure[]{
\includegraphics[scale=0.46]{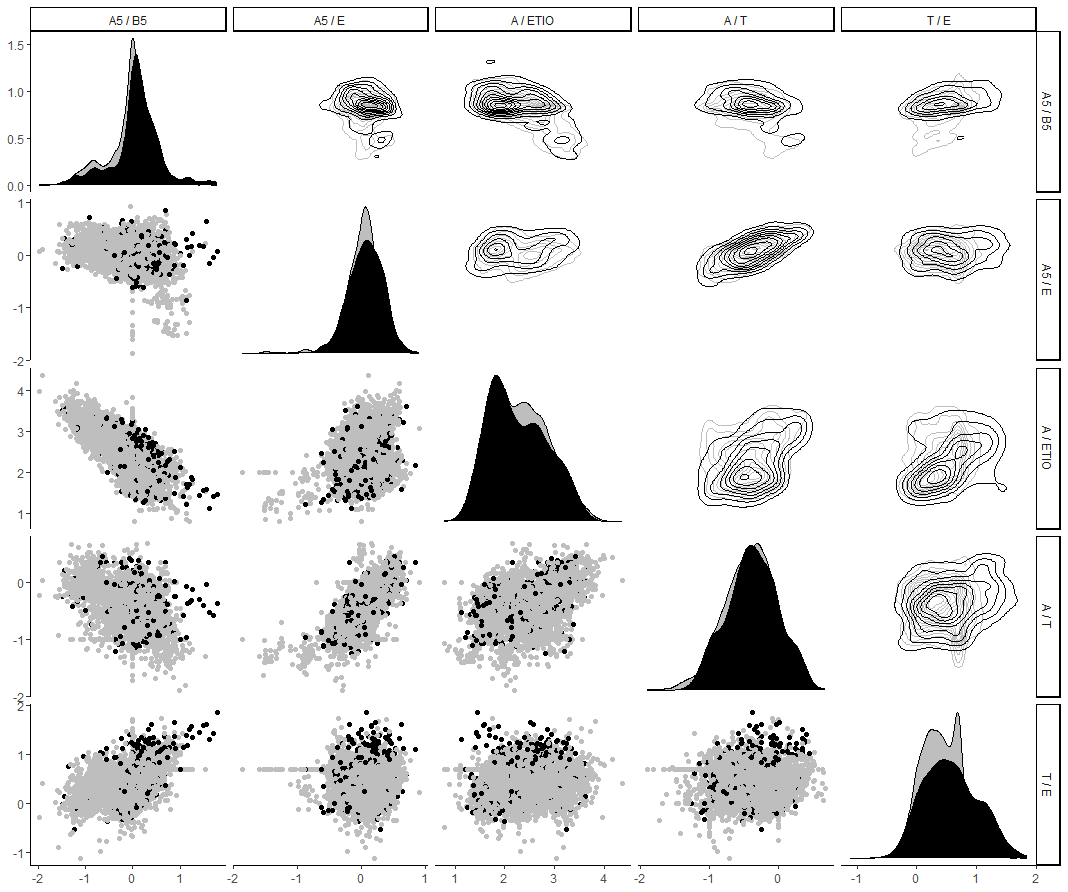}}  
\caption{Plots of normal (grey) vs non-normal (black) samples including the scatter, density and contour plots for (a) the six markers, and (b) their five ratios in the logarithmic scale.}
\label{fig:ggpairsplot_normalvsabnormal}
\end{figure}

\begin{figure}[!htbp]%
\centering
\subfigure[] {
\includegraphics[scale=0.43]{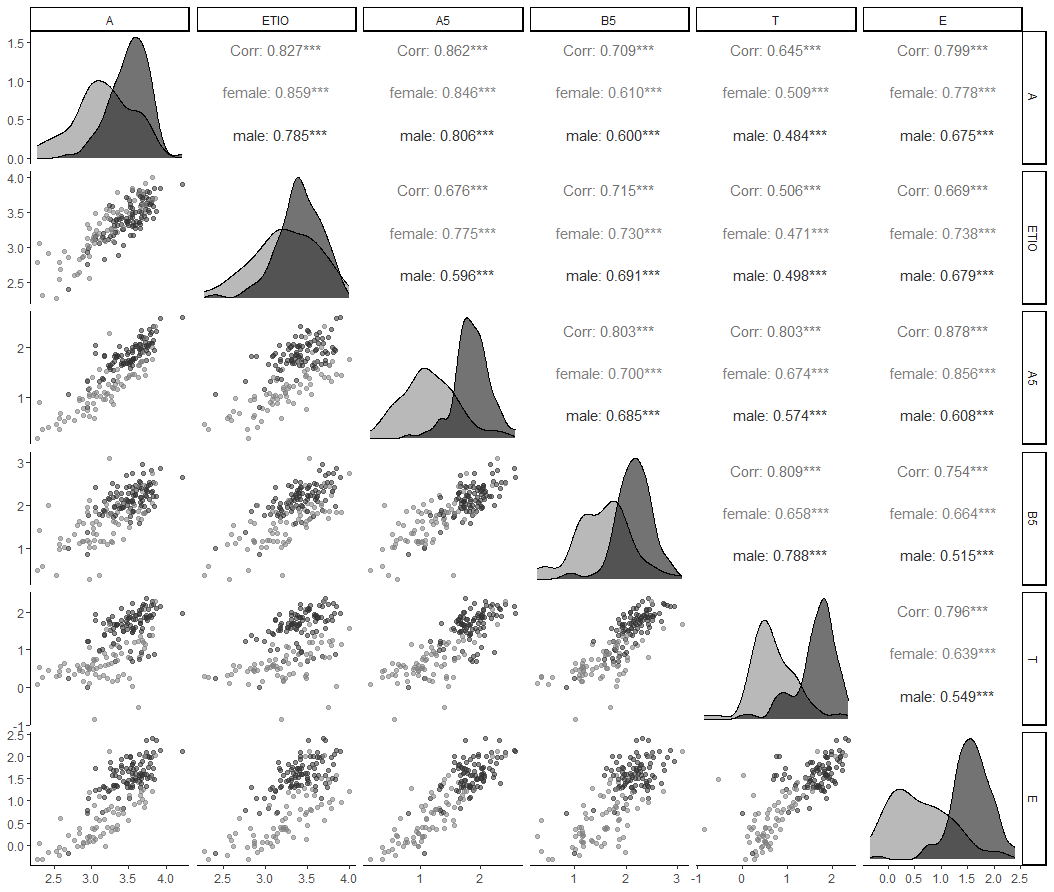}}
\qquad
\subfigure[]{
\includegraphics[scale=0.43]{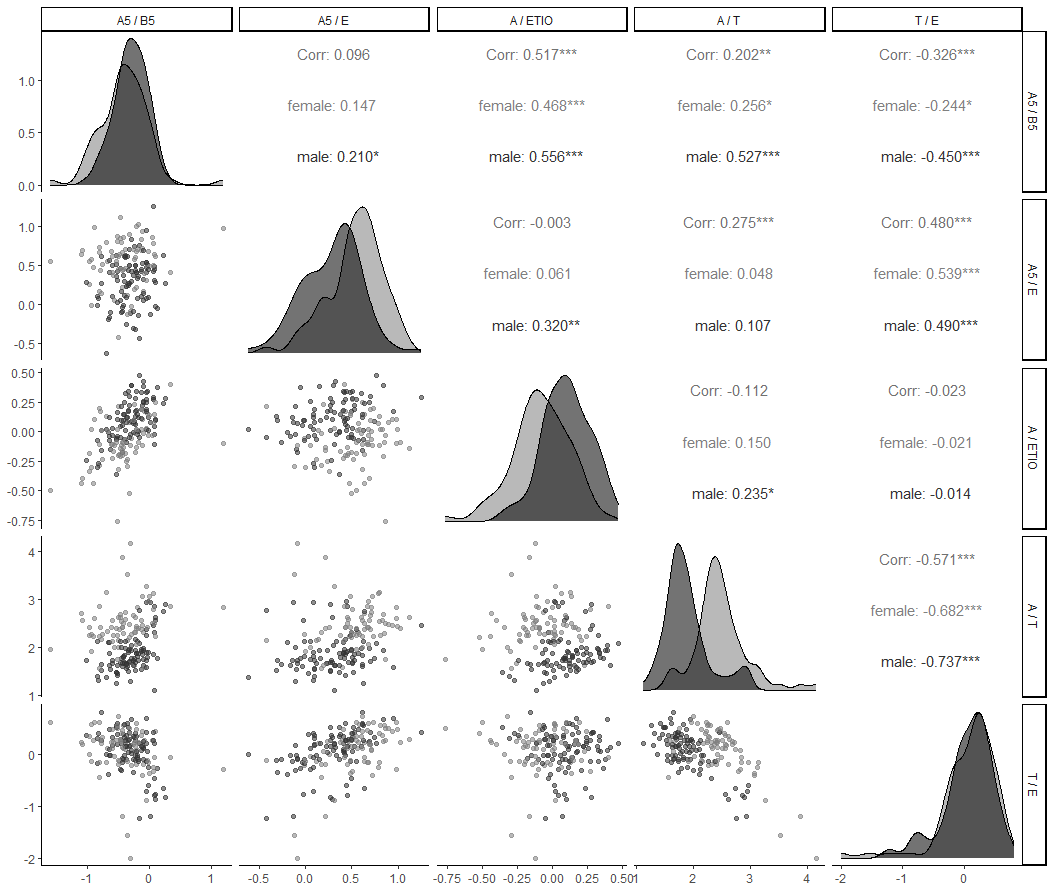}}  
\caption{Scatter and density plots by %gender
sex accompanied by the correlation coefficients for (a) the six markers, and (b) their five ratios in the logarithmic scale.}
\label{fig:ggpairsplot}
\end{figure}

\vspace{-0.5cm}
\begin{figure}[!htbp]%
\centering
\subfigure[] {
\includegraphics[scale=0.29]{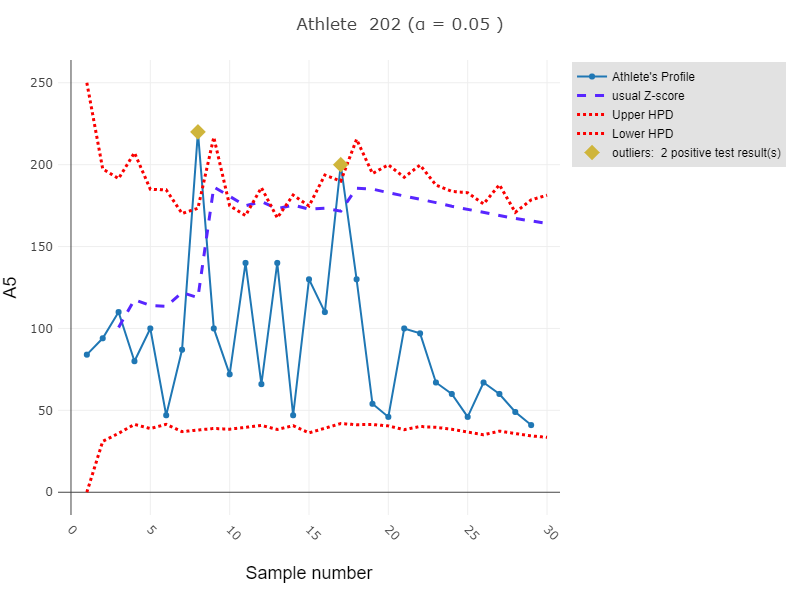}}
\subfigure[]{
\includegraphics[scale=0.29]{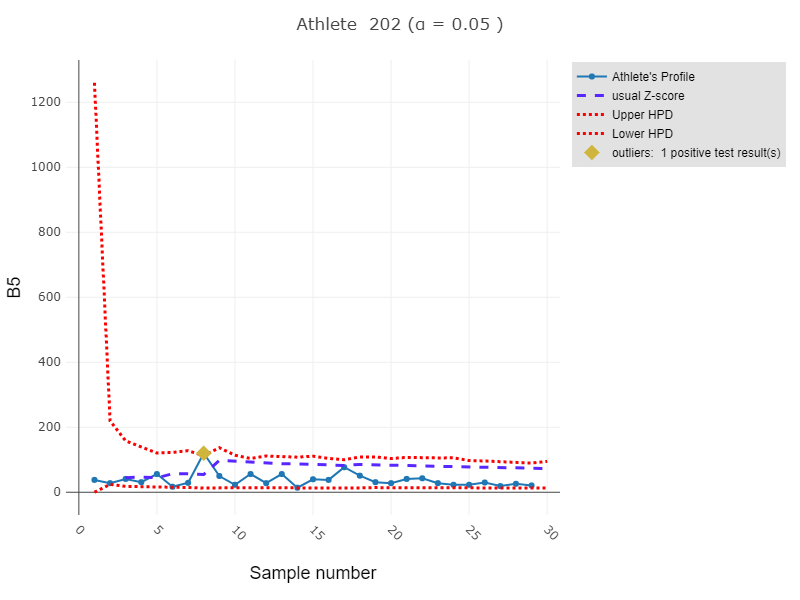}
}
\vspace{-0.3cm}
\subfigure[] {
\includegraphics[scale=0.29]{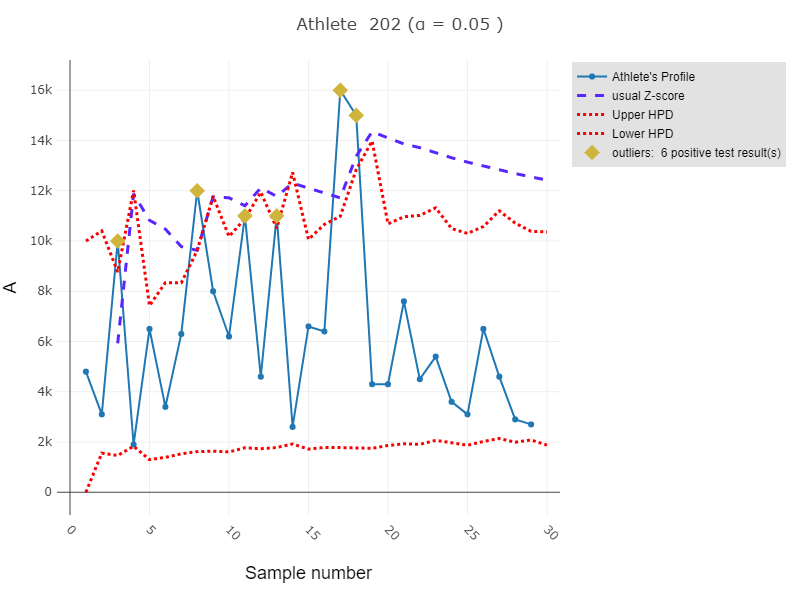}}
%\qquad
\subfigure[]{
\includegraphics[scale=0.29]{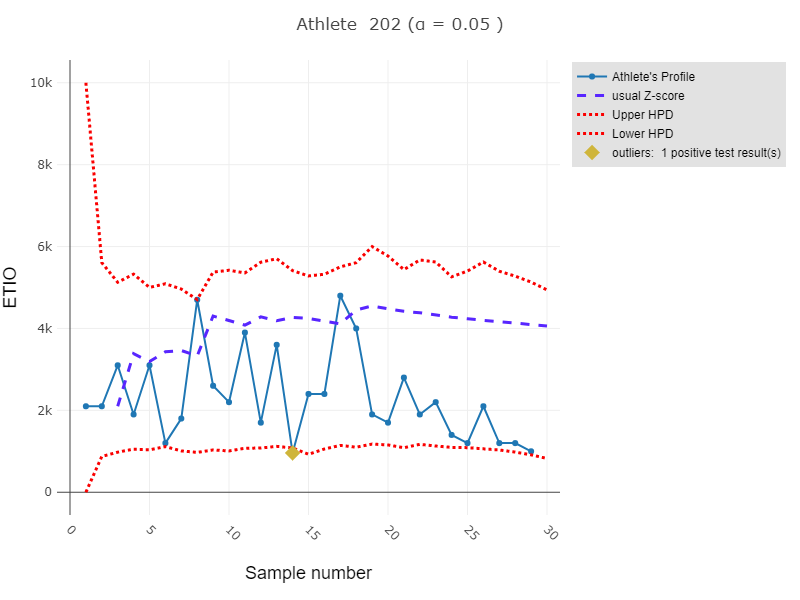}
}
\subfigure[]{
\includegraphics[scale=0.29]{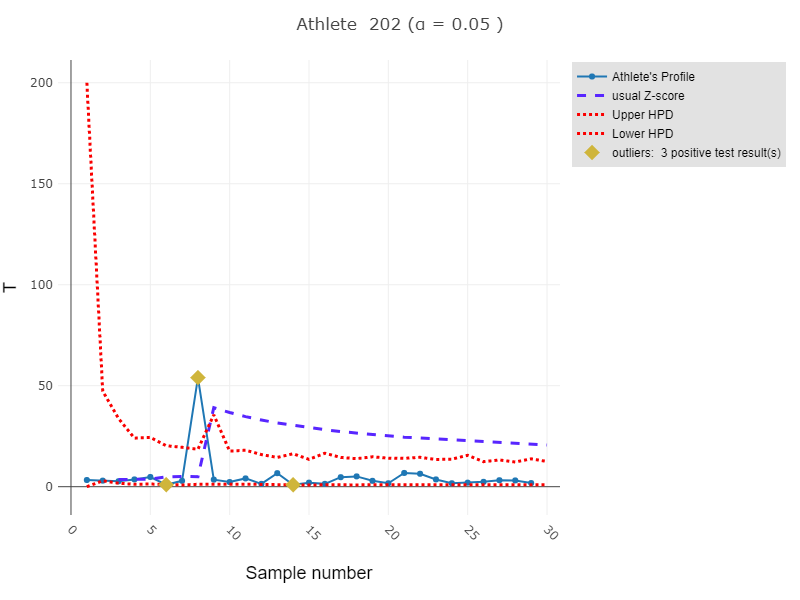}}
\subfigure[]{
\includegraphics[scale=0.29]{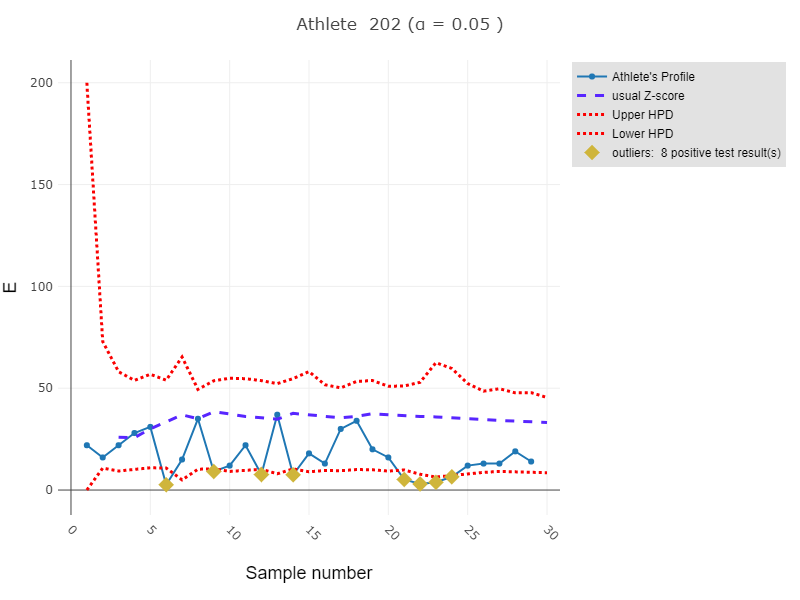}
}
\end{figure}

\begin{figure}[!htbp]%
\centering
\subfigure[] {
\includegraphics[scale=0.29]{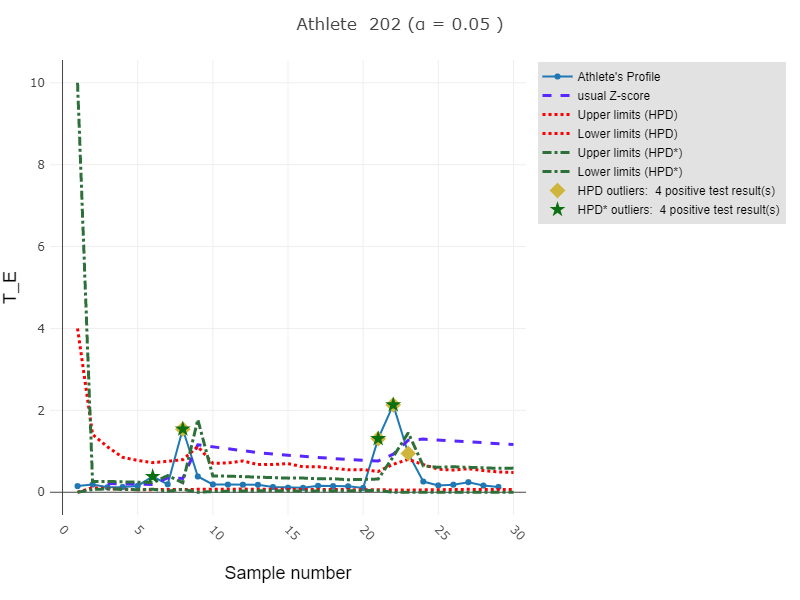}}
\subfigure[] {
\includegraphics[scale=0.29]{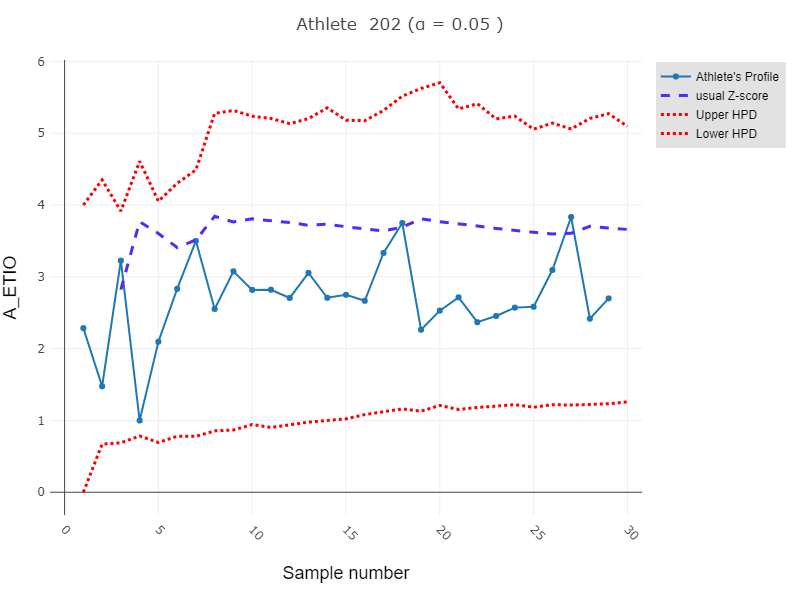}}
\vspace{-0.2cm}
\subfigure[]{
\includegraphics[scale=0.29]{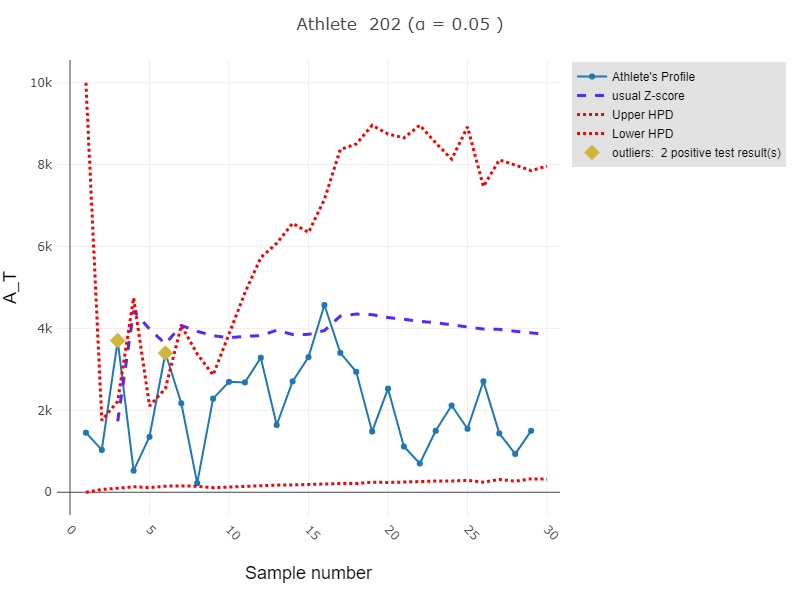}
}
\vspace{-0.2cm}
\subfigure[] {
\includegraphics[scale=0.29]{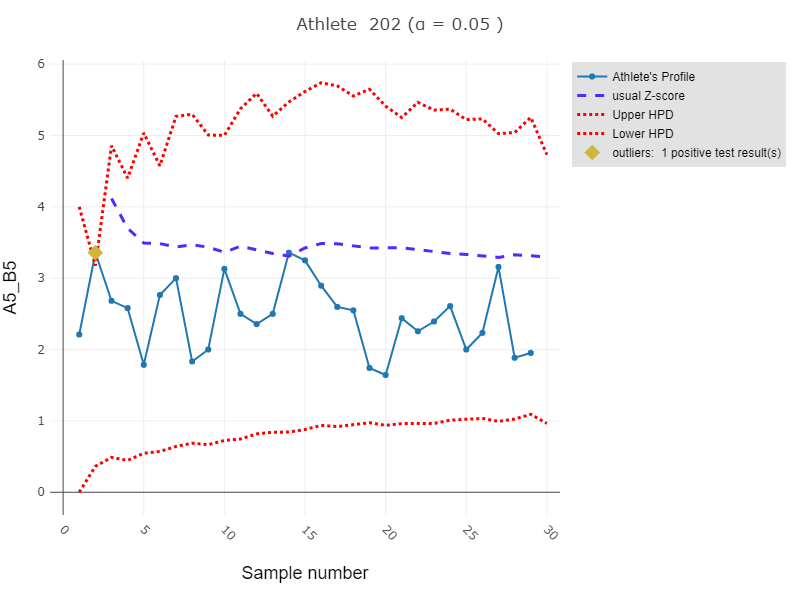}}
\vspace{-0.2cm}
\subfigure[]{
\includegraphics[scale=0.29]{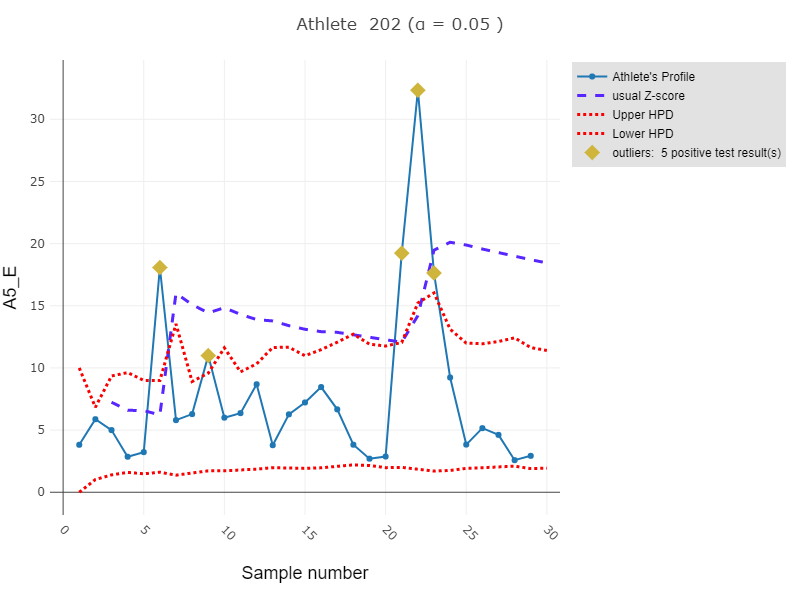}
}
\vspace{-0.2cm}
\caption{ (a-k) A series of 29 longitudinal values of the six EAAS and their five ratios (blue solid-dotted line) obtained from a doped athlete; upper and lower limits (red dotted lines) are calculated using the 95\% HPD intervals of the predictive distribution from the univariate Bayesian model; upper limits assuming a usual Z-score (purple dashed line); suggested abnormal values are denoted by the gold diamonds. (g) Upper and lower limits (green dashed-dotted lines) are calculated using the 95\% HPD interval of the predictive distribution from the T/E model of \citet{sottas2007bayesian}; suggested abnormal values based on the T/E model are denoted by the green stars.}
\label{fig:positiveprofiles} %male
\end{figure}

\newpage
\section{Tables}
\label{tables}

\setcounter{table}{0}
\renewcommand{\thetable}{B.\arabic{table}}
\renewcommand{\theHtable}{B.\thetable}

% table like table 1 of chan 2008
\vspace{-0.3cm}
\begin{table}[H]
\begin{center}
\small
\begin{tabular}{@{} *5l @{}}    \toprule
Trivial name            & Abbreviation & Systematic name & Formula  \\\midrule  
5$\alpha$-Adiol &      A5      &  5$\alpha$-Androstane-3$\alpha$,17$\beta$-diol  & $C_{19}H_{32}O_{2}$ \\
5$\beta$-Adiol &      B5      &  5$\beta$-Androstane-3$\alpha$,17$\beta$-diol  & $C_{19}H_{32}O_{2}$\\
Androsterone            &      A       & 3$\alpha$-Hydroxy-5$\alpha$-androstan-17-one   & $C_{19}H_{30}O_{2}$\\
%Dehydroepiandrosterone  &      DHEA    &  3$\beta$-Hydroxy-androst-5-en-17-one   & $C_{19}H_{28}O_{2}$\\ 
%5$\alpha$-Dihydrotestosterone  &      DHT     &    17$\beta$-Hydroxy-5$\alpha$-androstan-3-one  & $C_{19}H_{30}O_{2}$\\ 
Epitestosterone         &      E       &  17$\alpha$-Hydroxy-androst-4-en-3-one  & $C_{19}H_{28}O_{2}$\\ 
Etiocholanolone         &      ETIO    &  3$\alpha$-Hydroxy-5$\beta$-androstan-17-one  & $C_{19}H_{30}O_{2}$\\ 
Testosterone            &      T       &   17$\beta$-Hydroxy-androst-4-en-3-one    & $C_{19}H_{28}O_{2}$\\  
5$\alpha$-Adiol/5$\beta$-Adiol  &      A5/B5   &       & \\
5$\alpha$-Adiol/Epitestosterone                        &      A5/E    &       & \\
Androsterone/Etiocholanolone           &      A/ETIO  &       & \\
Androsterone/Testosterone              &      A/T     &       & \\
Testosterone/Epitestosterone           &      T/E     &       & \\\bottomrule
 \hline
\end{tabular}
\end{center} 
\vspace{-0.45cm}
\caption{\label{table:metabolites} Urinary steroid metabolites quantitated in this study.}
\end{table}

\begin{table}[H] 
\centering 
\footnotesize
\begin{tabular}{llllllll}
\\ \hline 
\hline \\  
 Target Metabolite  & Min  & Q1 & Mean & Median & Q3  & Max & SD \\ 
     &  (ng/mL) & (ng/mL) & (ng/mL)& (ng/mL) & (ng/mL) & (ng/mL) & (ng/mL) \\
\hline \\ 
A5       & $1.46$  & 14.49 & $61.7$    & $46.62$   & 84.56  & $388.14$   & $64.68$  \\ 
B5       & $1.88$  & 40.46 & $139.92$    & $89.14$   & 180.92 & $1,232.95$   & $164.27$  \\ 
A        & $187.7$ & 1,315.9& $2,997.3$ & $2,517.2$& 4,428.1 & $16,674.1$ & $2,169.41$ \\ 
E        & $0.47$   & 4.31 & $33.71$    & $20.52$      & 42.22& $252.96$    & $42.94$   \\ 
ETIO     & $187.5$ & 1,379.8 & $2,719.4$ & $2,337.9$& 3,614 & $9,819.9$ & $1,803.23$ \\ 
T        & $0.14$   & 4.17 & $39.37$    & $20.39$   & 61.43  & $229.03$    & $46.27$  \\ 
A5/B5    & $0.03$   & 0.3  & $0.64$     & $0.46$    & 0.74   & $15.57$     & $1.22$   \\ 
A5/E     & $0.24$   & 1.47  & $3.29$     & $2.73$    & 4.23   & $17.72$     & $2.54$   \\ 
A/ETIO   & $0.17$   & 0.78  & $1.17$     & $1.05$    & 1.47   & $3$     & $0.53$   \\ 
A/T      & $13.13$  & 53.97 & $374.49$    & $125.08$   & 282.92 & $14,154.24$   & $1,269.35$  \\ 
T/E      & $0.01$   & 0.76 & $1.63$     & $1.44$    & 2.19   & $6.48$     & $1.16$   \\ 
\hline \\  
\end{tabular} 
\caption{\label{table:164healthysummary} Descriptive summaries (minimum, %inter-quartile range; 
1st quartile; Q1, mean, median, 3rd quartile; Q3, maximum and standard deviation (SD)) of the metabolites and ratios of the baseline healthy population (91 men and 73 women). %The names of the target metabolites are abbreviated as presented in Table \ref{table:metabolites} in Appendix \ref{tables}.
}
\end{table}

\vspace{-0.5cm}
\begin{table}[H] 
\centering 
\footnotesize
\begin{tabular}{lllllll}
\hline 
\hline 
   & Min  & Q1 & Mean & Median & Q3  & Max  \\ 
 Target Metabolite    &  (ng/mL) & (ng/mL) & (ng/mL)& (ng/mL) & (ng/mL) & (ng/mL) \\
\hline \\ 
A5     & $1$    & 11    & $35.08$  & 25       & $49$  & 210      \\ 
B5     & $1$    & 34    & $96.25$  & $64$     & 120   & $910$  \\ 
A      & $100$  & 1,200 & $2,097$  & $1,800$  & 2,600 & $16,000$ \\ 
E      & $0.10$ & 4     & $17.48$  & $11$     & 26    & $130$    \\ 
ETIO   & $98$   & 1,100 & $1,863$  & $1,700$  & 2,400 & $7,200$ \\ 
T      & $0.10$ & 3.30  & $20.17$  & $9.60$   & 31    & $150$    \\ 
A5/B5  & $0.013$& 0.24  & $0.49$   & $0.40$   & 0.61  & $4.8$  \\ 
A5/E   & $0.13$ & 1.41  & $4.77$   & $2.54$   & 4.65  & $160$    \\ 
A/ETIO & $0.06$ & 0.80  & $1.21$   & $1.11$   & 1.48  & $8.16$   \\ 
A/T    & $6.25$ & 72.41 & $365.25$ & $159.46$ & 412.7 & $23,000$ \\ 
T/E    & $0.012$& 0.75  & $1.35$   & $1.0$   & 1.6   & $13$    \\ 
\hline \\  
\end{tabular} 
\vspace{-0.45cm}
\caption{\label{table:summary_long} Descriptive summaries (minimum, %inter-quartile range; 
1st quartile; Q1, mean, median, 3rd quartile; Q3, and maximum) of the metabolites and ratios of 100 athletes with normal samples.}
\end{table} 

\vspace{-0.2cm}
\begin{table}[H] 
\centering 
\footnotesize
\begin{tabular}{lllllll}
\hline 
\hline 
   & Min  & Q1 & Mean & Median & Q3  & Max  \\ 
Target Metabolite     &  (ng/mL) & (ng/mL) & (ng/mL)& (ng/mL) & (ng/mL) & (ng/mL) \\
\hline \\ 
A5     & $1.0$   & 14    & $36.61$    & 28.0     & $50$  & 250      \\ 
B5     & $3.3$   & 43    & 105.1      &$73$      & 140   & $1,400$  \\ 
A      & $100$   & 1,200 & 2,318      & $2,000$  & 3,000 & $13,000$ \\ 
E      & $0.10$  & 4.50  & 18.53      & $12.0$   & 27    & $160$    \\ 
ETIO   & $270$   & 1,300 & 2,199      & $1,900$  & 2,700 & $14,000$ \\ 
T      & $0.10$  & 3.3   & 19.44      & $9.25$   & 33    & $150$    \\ 
A5/B5  & $0.02$  & 0.19  & $0.48$     & $0.35$   & 0.65  & $4.6$  \\ 
A5/E   & $0.08$  & 1.42  & $3.38$     & $2.37$   & 4.6   & $53.33$  \\ 
A/ETIO & $0.014$ & 0.7   & $1.16$     & $1.04$   & 1.46  & $6.91$   \\ 
A/T    & $6.252$  & 67.66 & $454.63$   & $183.33$ & 525.65 & $9,200$ \\ 
T/E    & $0.01$ & 0.62  & 1.46       & $1$   & 1.64  & $35$  \\ 
\hline \\  
\end{tabular} 
\vspace{-0.45cm}
\caption{\label{table:summary_long2} Descriptive summaries (minimum, %inter-quartile range; 
1st quartile; Q1, mean, median, 3rd quartile; Q3, and maximum) of the metabolites and ratios of 100 athletes with atypical samples.}
\end{table} 

\vspace{-0.2cm}
\begin{table}[H] 
\centering 
\footnotesize
\begin{tabular}{lllllll}
\hline 
\hline 
   & Min  & Q1 & Mean & Median & Q3  & Max  \\ 
Target Metabolite     &  (ng/mL) & (ng/mL) & (ng/mL)& (ng/mL) & (ng/mL) & (ng/mL) \\
\hline \\ 
A5     & $3.7$   & 21    & $50.53$    & 39.5     & $65$  & 270      \\ 
B5     & $4.9$   & 24    & $82.69$     & $50$     & 97    & $2,200$  \\ 
A      & $210$   & 1,600 & $3,152$    & $2,400$  & 3,900 & $16,000$ \\ 
E      & $1$     & 5.90  & $14.41$    & $11$     & 19    & $120$    \\ 
ETIO   & $100$   & 932.5 & $1,747$    & $1,500$  & 2,175 & $11,000$ \\ 
T      & $0.50$  & 1     & $12.51$    & $3.50$   & 15    & $220$    \\ 
A5/B5  & $0.085$ & 0.47  & $0.76$     & $1$     & 1.42  & $4.2$  \\ 
A5/E   & $0.46$  & 1.83  & $5.59$     & $3.55$  & 5.91  & $72.22$  \\ 
A/ETIO & $0.50$ & 1.17  & $1.92$     & $1.69$   & 2.53  & $5.08$   \\ 
A/T    & $13.18$  & 177.55& $1,262.99$ & $632.46$ & 1,800 & $14,828$ \\ 
T/E    & $0.03$  & 0.15  & $1.58$     & $0.34$   & 1.3  & $61.11$  \\ 
\hline \\  
\end{tabular} 
\vspace{-0.45cm}
\caption{\label{table:summary_long3} Descriptive summaries (minimum, %inter-quartile range; 
1st quartile; Q1, mean, median, 3rd quartile; Q3, and maximum) of the metabolites and ratios of 29 athletes with abnormal samples.}
\end{table}

\begin{table}[H] 
\centering 
\footnotesize
\begin{tabular}{lllll}
\\ \hline 
\hline \\  
 Target Metabolite  & Max (M)  & Max (F) & WADA TD 2021 (M)  &  WADA TD 2021 (F)  \\ 
   &  (ng/mL)    & (ng/mL)      &  (ng/mL)  &  (ng/mL)     \\
\hline \\ 
A5       &    652   & 263    & 250     & 150    \\ 
B5       &   1,260  & 471    &         &        \\ 
A        &  20,700  & 17,500 & 10,000  & 10,000 \\ 
E        &    391   & 51.9   & 200     &  50    \\ 
ETIO     & 11,400   & 9,030  & 10,000  & 10,000 \\ 
T        &   249    & 219    & 200     &  50    \\ 
A/ETIO   &          &        &  4      &   4    \\
T/E      &          &        &  4      &   4    \\ 
\hline \\
Target Metabolite & IUL (M)   &  IUL (F)  & & \\
                  &  (ng/mL) &  (ng/mL) & & \\
\hline \\
A5/B5    &     4        &   4      &       &       \\ 
A5/E     &     10       &   10     &       &       \\ 
A/T      &     10000    &   10000  &       &       \\ 
\hline \\  
\end{tabular} 
\caption{\label{table:WADAreflimits} Maximum EAAS values and their ratios measured in a Caucasian population consisting of 2,027 male (M) and 1,004 female (F) athletes \cite{van2010reference} along with the available WADA's threshold limits by sex \cite{WADA2018techdoc, WADA2021TDAPMU}. The initial upper limits (IUL) for ratios, in cases where there is no available population-specific information, are determined from the upper limit of the inter-quartile range (Q3) observed in the corresponding ratios within the cross-sectional dataset.}
\end{table}

\end{document}